\documentclass[preprint,12pt]{elsarticle}

\usepackage{graphicx}
\usepackage{placeins}
\usepackage{amssymb}

\usepackage{lineno}
\usepackage{gensymb}

\usepackage{subfigure}

\begin{document}

\begin{frontmatter}


\title{Mechanics of Topologically Interlocked Material Systems under Point Load: Archimedean and Laves Tiling}



\author{Andrew Williams, Thomas Siegmund(1)}

\address{School of Mechanical Engineering, Purdue University, West Lafayette, IN 47907, USA;
(1) Corresponding Author, siegmund at purdue.edu}

\begin{abstract}
Topologically interlocked material systems are two-dimensional assemblies of unit elements from which no element can be removed from the assembly without disassembly of the entire system.
Consequently, such tile assemblies are able to carry transverse mechanical loads.
Archimedean and Laves tilings are investigated as templates for the material system architecture.
It is demonstrated under point loads that the architecture significantly affects the force-deflection response.
Stiffness, load carrying capacity and toughness varied by a factor of at least three from the system with the poorest performance to the system with the best performance.
Across all architectures stiffness, strength and toughness are found to be strongly and linearly correlated.
Architecture characterizing parameters and their relationship to the mechanical behavior are investigated. 
It is shown that the measure of the smallest tile area in an assembly provides the best predictor of mechanical behavior.
With small tiles present in the assembly the contact force network structure is well developed and the internal load path is channeled through these stiffest components of the assembly.
\end{abstract}


\end{frontmatter}


\section{Introduction}
\label{S:1}

Plates are ubiquitous two-dimensional structural units able to carry transverse loads. Commonly, plates are monolithic \cite{lowe2012basic}, but plates-type structures can also be assembled from topologically interlocked unit elements in the form of convex polyhedra. Planar assemblies of convex polyhedra were considered as early as in the 17th century \cite{gallon1777machines}. A renewed interest in such structures occurred recently in the civil engineering \cite{Glickman1984,rippmann2013rethinking,fallacara2018unfinished,Vella2016GeometricVault,tessmann2012topological,tessmann2019geometry,dyskin2005principle} and materials engineering context \cite{dyskin2001toughening,Dyskin2001AElements,barthelat2015architectured,ries2013,Siegmund2016ManufactureAssemblies,wang2019design,estrin2003topological}. 

In such assemblies individual building blocks (or tiles) are shaped and arranged in the assembly such that no building block can be removed without the disassembly of the overall system. When considering such systems in the context of material design \cite{ashby2005hybrids,ashby2011hybrid,Estrin2011TopologicalConcept} they provide a unique method to expand the material property space and for quasi-static loading has been demonstrated to enable the transformation of a brittle response of a monolithic plate made of brittle materials (such as ceramics, glasses or brittle polymers) into a quasi-ductile response in the assembled plate-type structure \cite{mather2012structural,Valashani2015ANacre,Dyskin2003FractureContacts}. Moreover, \cite{Mirkhalaf2018SimultaneousCeramics} demonstrated that for certain classes of solid-architecture combination a simultaneous improvement of strength and toughness of the assembled plate relative to the monolithic plate is possible. Such favourable mechanical performance of the assembled plate structures also were found to extend to impact loading by altering the relationship between impact velocity and residual velocity \cite{Feng2015ImpactAssemblies} and increased impact energy absorption capacity \cite{javan2016design,javan2017impact,RezaeeJavan2018ImpactStudy}.
In addition, assembled plate structures can serve as the template for the implementation of adaptive structural configurations \cite{Khandelwal2015AdaptiveSystems,molotnikov2015design} to control system stiffness, strength and toughness.

In such prior work interlocked assemblies of building blocks have commonly been considered from the viewpoint of assemblies of all identical building blocks \cite{Dyskin2003TopologicalStructures}. Such a viewpoint is limiting on what types of architectures can be obtained. The material architecture can be expanded when the starting point for the construction of the interlocked material system is an underlying grid instead of the particles \cite{Weizmann2017TopologicalFloors, Weizmann2016TopologicalFloors, Weizmann2015TopologicalDesign, Bejarano2019AConfigurations,piekarskinew,Rippmann2018ComputationalVaults}. The construction of topologically interlocked material systems emerging from underlying grid systems is best placed into the context of the theory of tessellations \cite{Grunbaum2016TilingsPatterns} as such an approach provides ordering principles for the architectures of concern. The rules set forth in \cite{Kanel-Belov2010InterlockingSolids} are then applicable to realize the interlocking building blocks related to a tessellation pattern.

The present study is connected to a background of prior work on the mechanics of plate-type topologically interlocked assemblies. Prior work on the mechanics of flat vaults \cite{Brocato2012ABond, Brocato2018, fantin2019resistance,pfeiffer2019topological}  has focused on the stability of such systems under gravity loads while a second body of work has considered applied displacement loads \cite{Khandelwal2013ScalingLoading,Khandelwal2012TransverseMaterials,short2019scaling,Mirkhalaf2018SimultaneousCeramics,dyskin2019topological}. What has emerged is that an understanding of the load-deformation response plate-type topologically interlocked assemblies clearly cannot be conducted within the framework of monolithic plates, and that the assembly architecture shall be an integral part of the description of the respective mechanical response.

There has been an absence of systematic investigations into the mechanical behavior of architectured plate systems constructed on the basis of underlying grid systems (tessellations). This study seeks to fill this gap with the ultimate objective to determine how the mechanical response of architectured plates relates to the underlying tessellation patterns. All possible Archimedean and Laves tilings are investigated. Cross-property relationships between stiffness, strength and toughness \cite{Ashby2010MaterialCharts} are determined as relationships between the plate architecture and the plate mechanical response characteristics.

\section{Methods}
\label{S:2}
\subsection{Interlocking Assemblies}
The midplane cross section of a topologically interlocked material (TIM) system is a 2D tiling, and this tiling is considered as the basis for the creation of the TIM system \cite{Kanel-Belov2010InterlockingSolids}.
TIM assemblies are considered as assemblies of blocks (polyhedra) which have center sections conforming to the tilings. Building blocks are constructed from the tiles of the tessellation by first projecting planes from each edge of the tile at alternating angles $\theta$ from the normal. In the following the construction principle is reviewed. Without loss in generality the principles are depicted for a square tiling. Code for the generation of the respective geometries is available~\cite{williams-python}.

\newcommand{\figWidthTrunc}{0.2}
\begin{figure}[!b]
    \centering
    \subfigure[]{
        \includegraphics[width=\figWidthTrunc\textwidth]{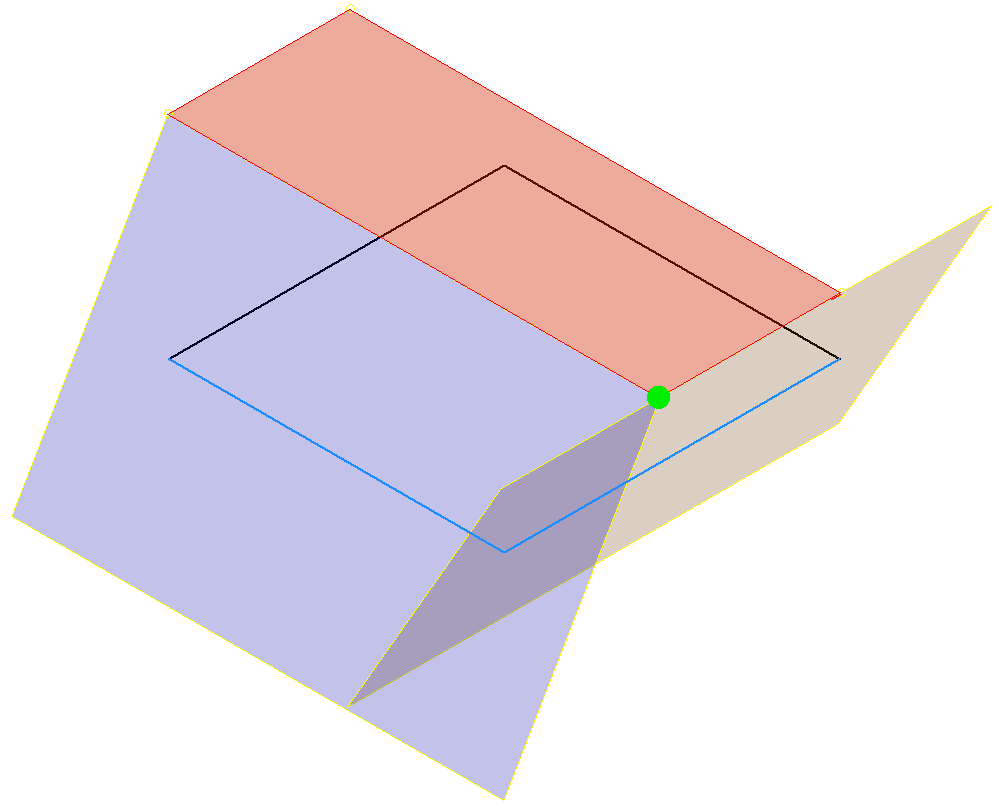}
        \label{trunc_top1}}
    \subfigure[]{
        \includegraphics[width=\figWidthTrunc\textwidth]{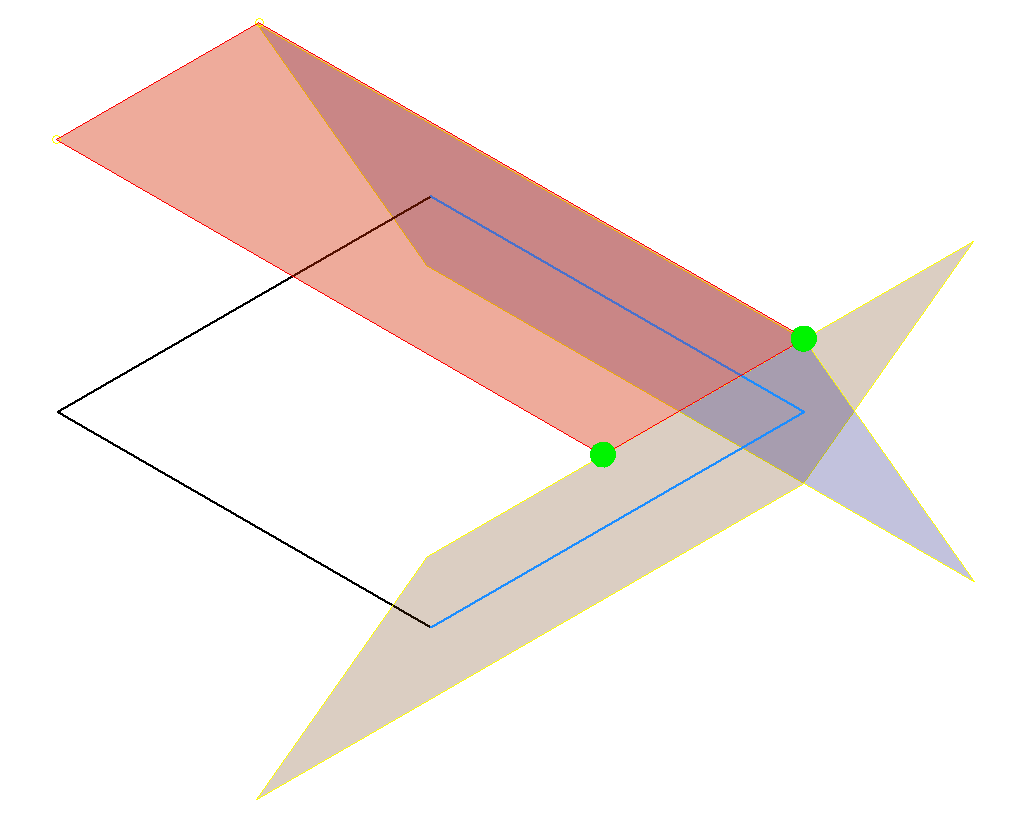}
        \label{trunc_top2}}
    \subfigure[]{
        \includegraphics[width=\figWidthTrunc\textwidth]{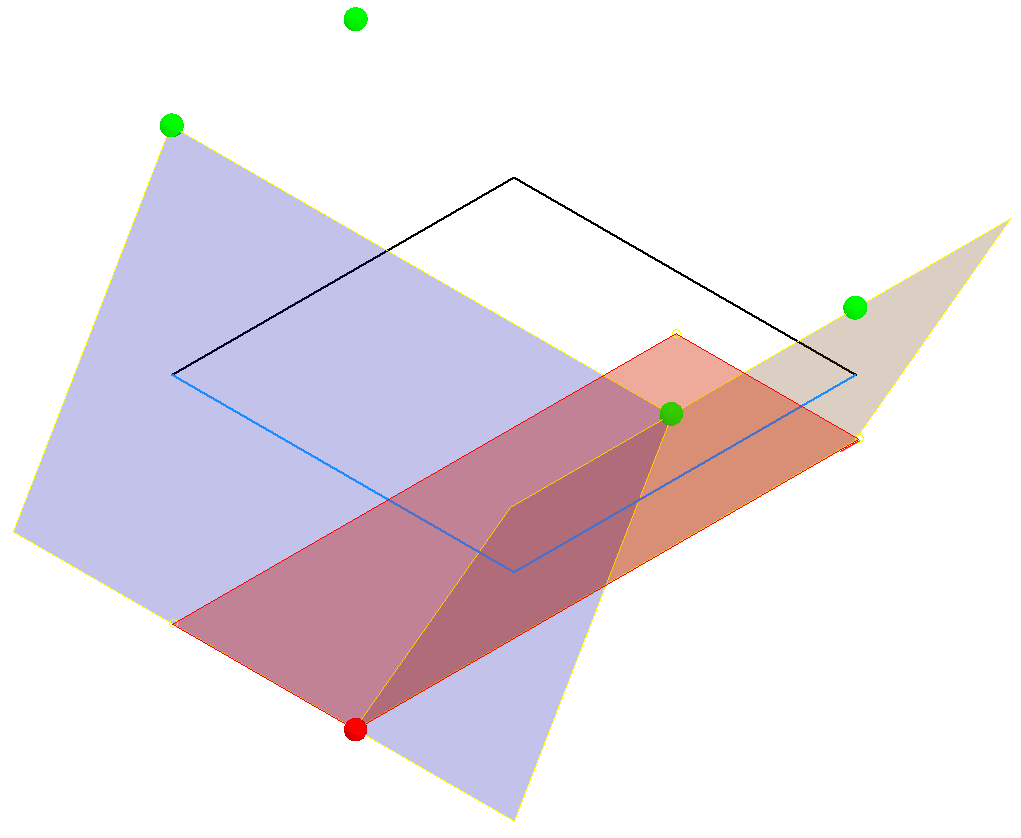}
        \label{trunc_bot1}}
    \subfigure[]{
        \includegraphics[width=\figWidthTrunc\textwidth]{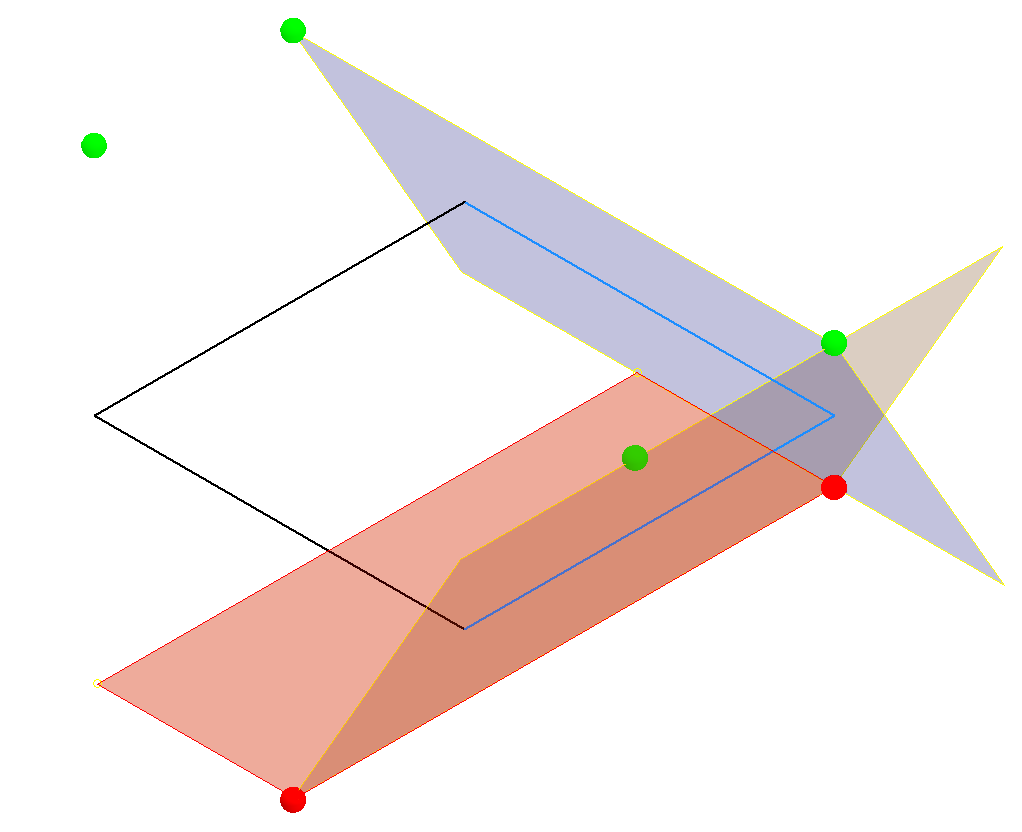}
        \label{trunc_bot2}}
    \subfigure[]{
        \includegraphics[width=\figWidthTrunc\textwidth]{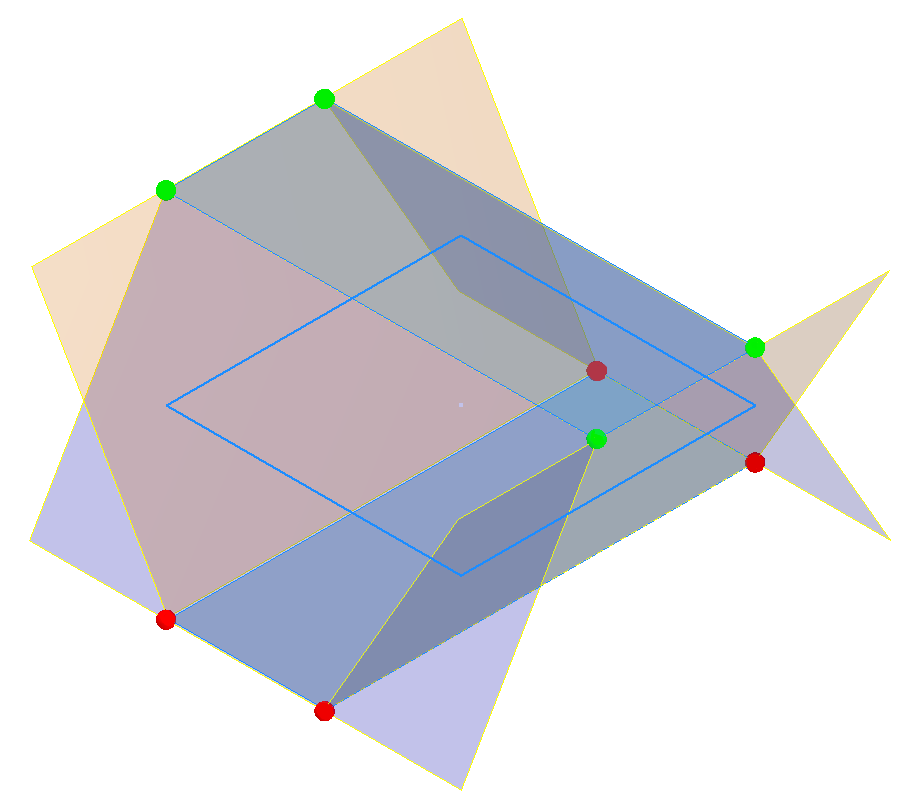}
        \label{trunc_all}}
    \subfigure[]{
        \includegraphics[width=0.16\textwidth]{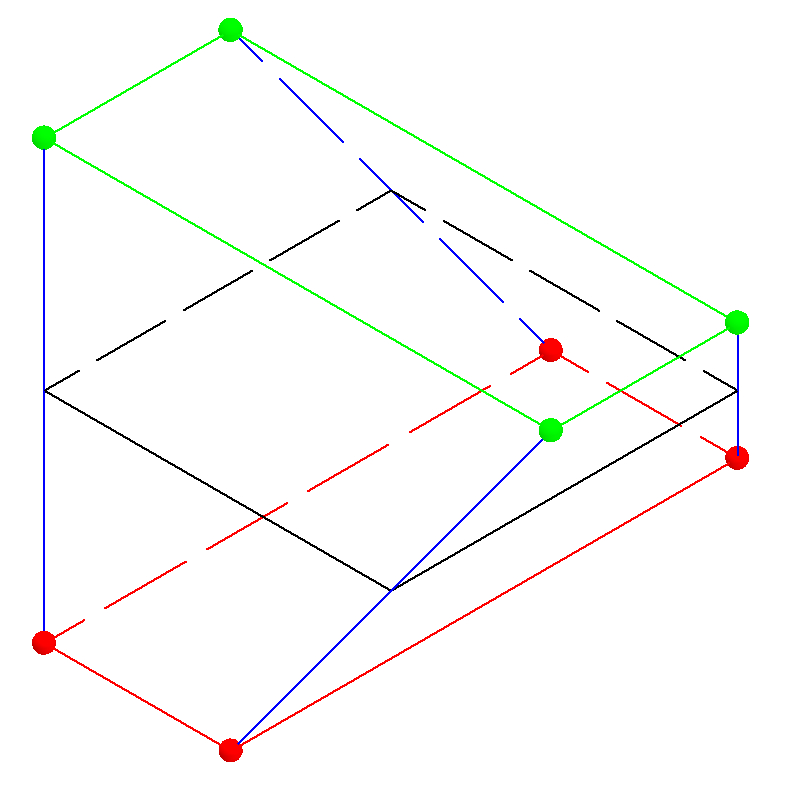}
        \label{trunc_wire}}
    \subfigure[]{
        \includegraphics[width=0.16\textwidth]{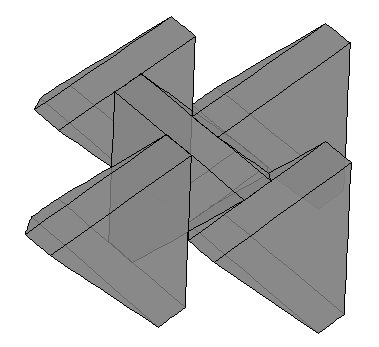}
        \label{assembly}}
        \caption{Truncated block construction from a square tile. (a) The intersection of two edge planes and the top plane defines the first top vertex. (b) The intersection of the next two edge planes and the top plane defines the second top vertex. (c) The intersection of two edge planes and the bottom plane defines the first bottom vertex. (d) The intersection of the next two edge planes and the bottom plane defines the next bottom vertex. (e) All planes and all vertices. (f) Wire frame of the resulting block. (g) Assembly of building blocks.}
    \label{trunc}
\end{figure}

The magnitude of the edge projection angle $\theta$ is a fixed value, but its direction will alternate between angling toward the tile center and away from the tile center for each edge.
The projection angle for all configurations in this study was $\theta = 17$\degree.
Within an assembly, the blocks must be oriented such that their edge projection angles are complimentary; if the edge of one tile is angled toward the tile center, the abutting edge of the adjacent tile must be angled away from its center.
Once the projection angles are specified, the vertices of the block can be determined.
Each block formed from an $n$-sided tile will have $n$ vertices, and if the tile is a regular polygon, a uniform antiprism block will be formed. Blocks constructed from tiles of different sizes and shapes naturally have differing overall dimensions. In order to control the aspect ratios of the TIM systems it is necessary to truncate the polyhedra to possess a common top and bottom plane in an assembly. Two additional planes must be defined parallel, at distance $H_0$ and equidistant from the tiling plane. Each building block (i.e. the trucated polyhedra) formed from an $n$-sided tile now possesses $2n$ vertices.
Every set of planes projecting from two consecutive tile edges will yield two vertices, one by computing their intersections with the top plane, Fig.~\ref{trunc}(a-b), and the other by computing their intersection with the bottom plane, Fig.~\ref{trunc}(c-d).
Computing the intersection of all sets of planes projecting from two consecutive edges and the top or bottom planes will locate all the vertices of the block, Fig.~\ref{trunc}(e).
Edges are then drawn between the vertices to construct the block, Fig.~\ref{trunc}(f). In the interlocking assembly, Fig.~\ref{trunc}(g), neighboring blocks impose constraints on each other such that assembly is load carrying.

\subsection{Tile Spaces}
The Archimedean and the Laves tilings are considered \cite{Grunbaum2016TilingsPatterns}. These tile sets are duals to each other.
Archimedean tilings consist of regular polygons only and possess one type of vertex. Laves tilings are defined as having an equal angular spacing of all edges at any vertex \cite{Grunbaum2016TilingsPatterns}. There are 11 Archimedean and 11 Laves tilings. Their structure is described by the naming convention of \cite{Grunbaum2016TilingsPatterns}. Integer numbers with exponents separated by periods and contained within parenthesis describe the common vertex at all tile intersections such that each integer represents the number of sides of a tile that shares the vertex, and the exponent is the number of that type of tile that shares the vertex.

\begin{figure}[!b]
    \setcounter{subfigure}{0} 
    \centering
    \subfigure[]{
        \label{arch_tiles}
        \includegraphics[scale=0.4]{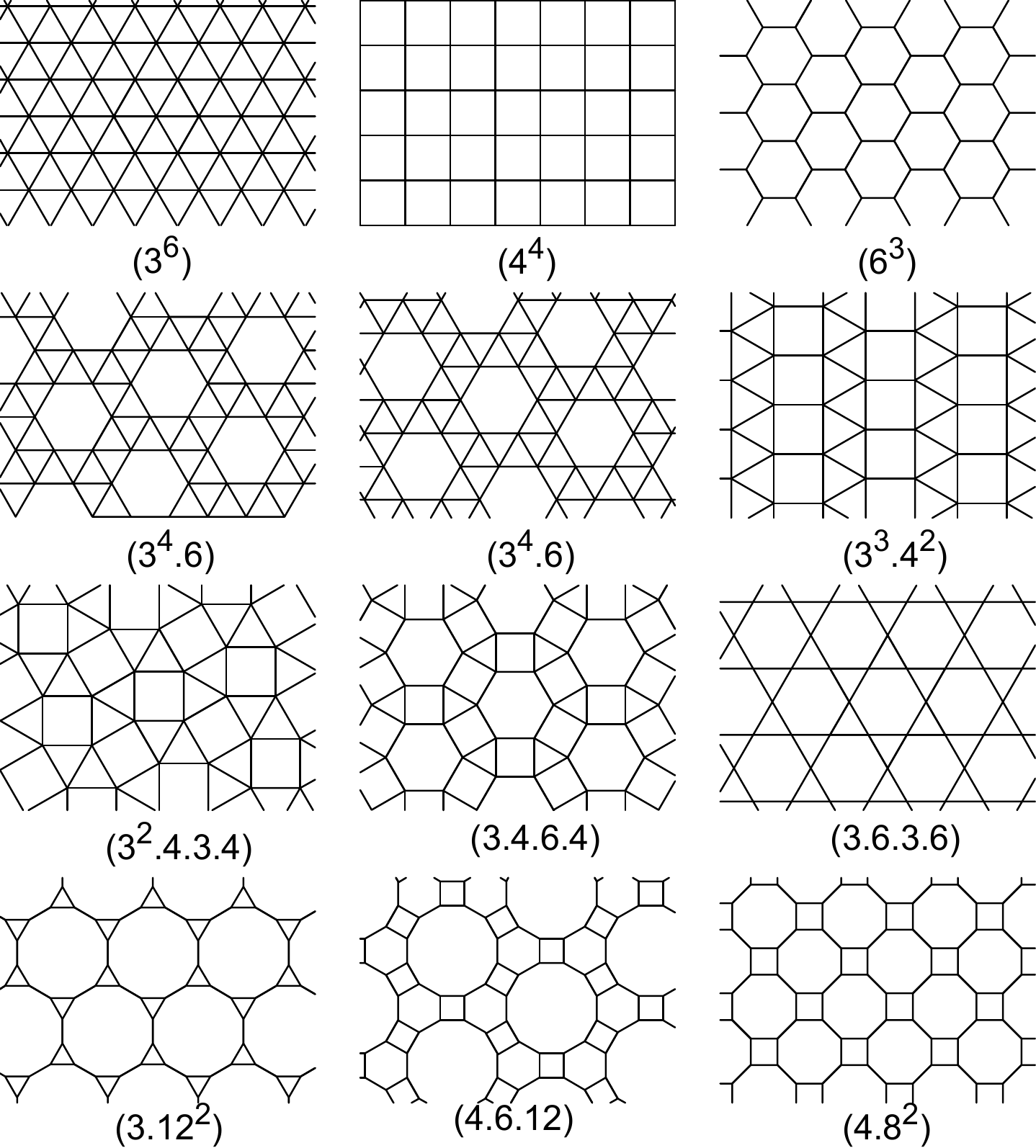}}
    \subfigure[]{
        \label{laves_tiles}
        \includegraphics[scale=0.4]{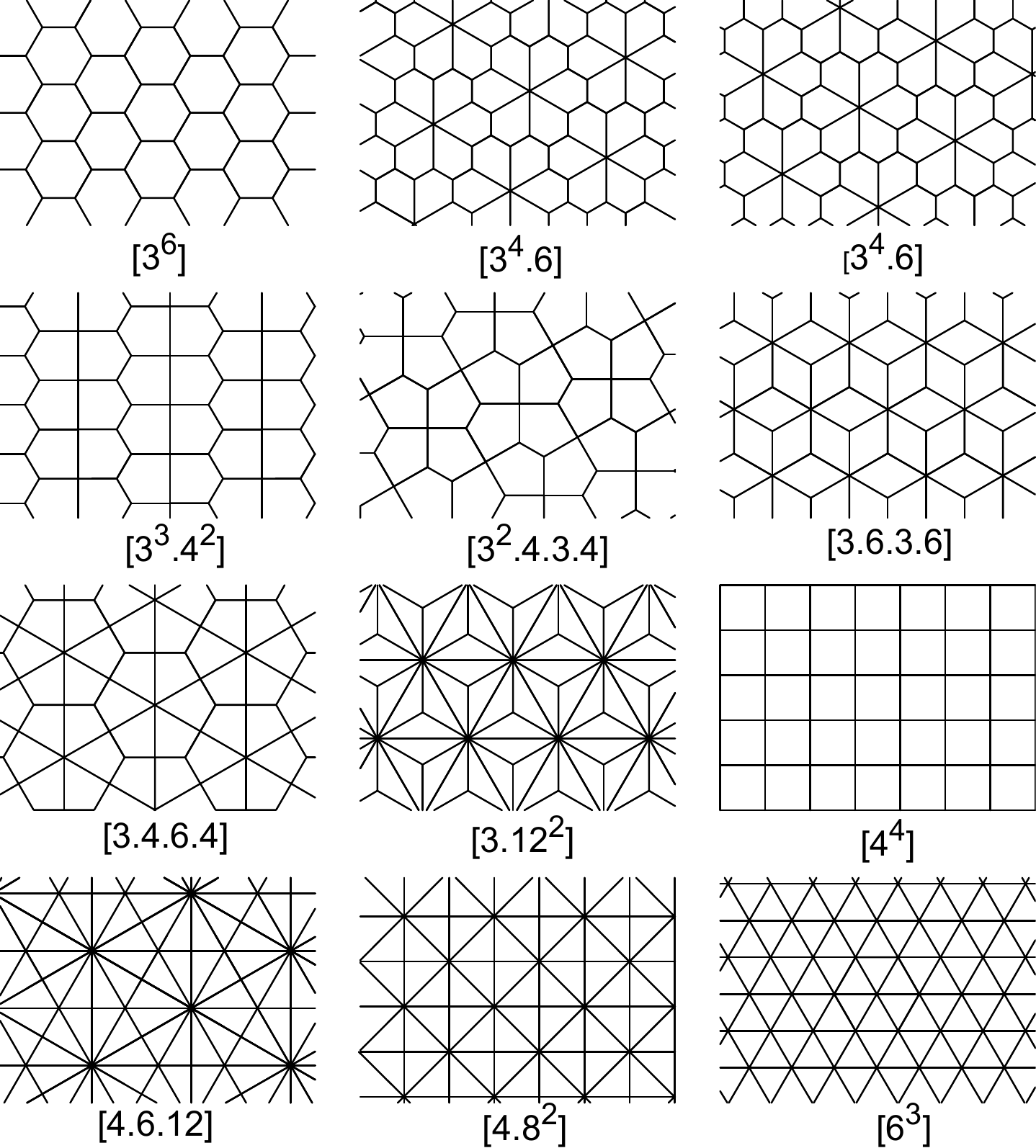}}
    \subfigure[]{
        \label{tile_set}
        \includegraphics[scale=0.5]{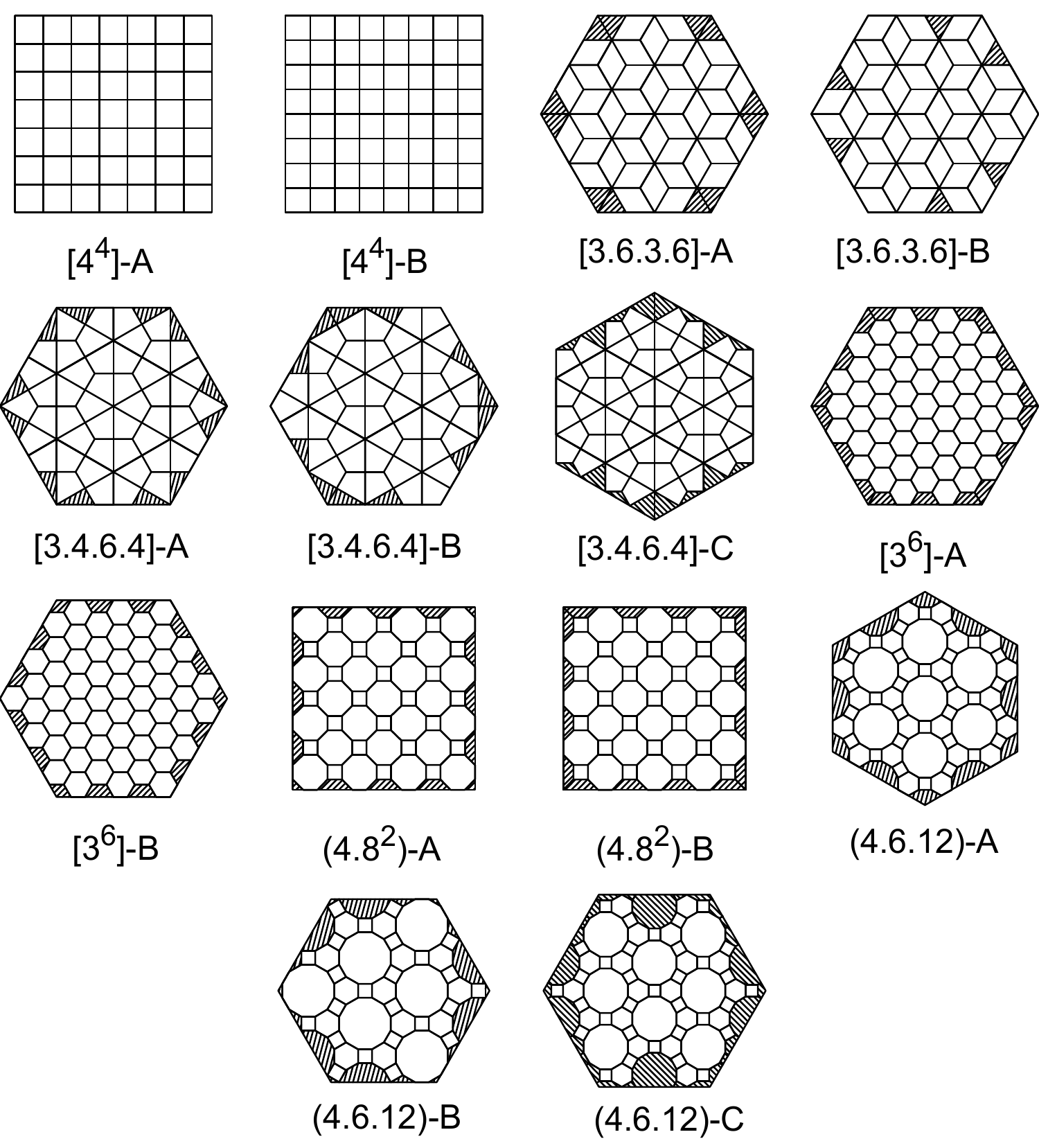}}
    \caption{(a) The 11 distinct Archimedean tilings. The $(3^4.6)$ tiling occurs in two forms, both are shown here. (b) The 11 distinct Laves tilings. The $[3^4.6]$ tiling occurs in left-handed and right-handed forms, both are shown here. (c) The set of bounded tilings considered in this work.}
    \label{fig:tilings}
\end{figure}

The Archimedean tilings are shown in Fig.~\ref{fig:tilings}(a). In a TIM system, the sides of each block must alternate between sloping toward and away from the normal to the plane of tessellation. Therefore, all tiles are required to possess an even number of sides. This restriction eliminates the $(3^6)$, $(3^4.6)$, $(3^3.4^2)$, $(3^2.4.3.4)$, (3.4.6.4), (3.6.3.6), and $(3.12^2)$ tilings for consideration as a TIM system.
The remaining tilings from which TIM systems can be constructed are $(4^4)$, $(6^3)$, (4.6.12), and $(4.8^2)$.

The Laves tilings are shown in Fig.~\ref{fig:tilings}(b). Again, TIM systems can only be constructed from a subset of the Laves tilings. The necessity for tiles with an even number of sides when constructing a TIM system eliminates the $[3^4.6]$, $[3^3.4^2]$, $[3^2.4.3.4]$, $[3.12^2]$, [4.6.12], $[4.8^2]$, and $[6^3]$ tilings.
The remaining tilings are the $[3^6]$, $[3.6.3.6]$, $[3.4.6.4]$, and $[4^4]$ tilings.
The $[4^4]$ and $[3^6]$ tilings are regular tilings and are equivalent to the $(4^4)$ and $(6^3)$ regular Archimedean tilings.
Therefore, only the $[3.6.3.6]$ and $[3.4.6.4]$ tilings are added beyond those from the Archimedean tilings.

In summary, the tilings suitable to TIM system construction are the $(4^4)$ (or $[4^4]$), $[3.6.3.6]$, $[3.4.6.4]$, $(6^3)$ (or$[3^6]$), $(4.8^2)$, and $(4.6.12)$ tilings.
The Laves notation was chosen to denote the $[4^4]$ and $[3^6]$ tilings instead of the Archimedean notation of $(4^4)$ and $(6^3)$ because these tilings are more similar to the other Laves tilings than to the other Archimedean tilings in this study.
The $[4^4]$, $[3.6.3.6]$, $[3.4.6.4]$, and $[3^6]$ tilings each consist of a single tile, whereas the $(4.8^2)$ tiling consists of two different tiles and the (4.6.12) tiling consists of three different tiles.

By their definition, tilings expand infinitely within a plane, yet here finite size assemblies are considered. 
Boundaries in the form of a regular polygon are defined for each tiling, such that the tiling was radially symmetric about its center point within the boundary. Squares or hexagons meet this criteria depending on the tiling but it is generally not possible to draw such a boundary without crossing any of the tiles. In such cases, any tiles that were intersected by the border became part of the border.
Furthermore, there are multiple possible center points for each tiling, such as centering the border around different vertices or around the centroid of different tiles. These various boundaries are referred to as A, B, and C variants of a given tiling. 

\begin{table}[ht]
    \centering
    \caption{Number of tiles and edges lengths for the set of bounded tilings in this work.}
    \begin{tabular}{|c|c|c|c|}
        \hline
        Tiling & Tiles & Edge 1 [mm] & Edge 2 [mm]\\[2pt]
        \hline
        $[4^4]$-A & 49 & 29.7 & -\\
        $[4^4]$-B & 64 & 26.0 & -\\
        $[3.6.3.6]$-A & 42 & 30.0 & -\\
        $[3.6.3.6]$-B & 46 & 30.0 & -\\
        $[3.4.6.4]$-A & 48 & 20.0 & 34.6\\
        $[3.4.6.4]$-B & 48 & 20.0 & 34.6\\
        $[3.4.6.4]$-C & 64 & 17.3 & 30.0\\
        $[3^6]$-A & 55 & 15.0 & -\\
       $[3^6]$-B & 57 & 15.0 & -\\
        $(4.8^2)$-A & 49 & 12.2 & -\\
        $(4.8^2)$-B & 49 & 12.2 & -\\
        (4.6.12)-A & 61 & 12.7 & -\\
        (4.6.12)-B & 43 & 14.6 & -\\
        (4.6.12)-C & 61 & 12.6 & -\\
        \hline
    \end{tabular}
    \label{tab:tilings}
\end{table}

All tilings were made into plate equivalent structures of fixed thickness value and aspect ratio.  The thickness of all assemblies was set to $H_0$=10.0 mm.
Square and hexagonal shaped assemblies are identical in that $L_0$ is the radius of the circle inscribed into the square or hexagon, Fig.~\ref{fig:tilings}(c). Prior work \cite{Khandelwal2013ScalingLoading} has shown that a minimum of 7 unit blocks per edge of the assembly is required to create TIM systems suitable for investigation.  The value of the in-plane dimension $L_0$ was derived for the geometric constraints imposed by the (4.6.12)-C assembly. This assembly, by nature of the combination of large and small building blocks imposes an upper limit on $L_0$. Geometric constraints for the (4.6.12)-C assembly with 61 blocks lead to an assembly having the ratio $L_0/H_0=10.39$. This value of $L_0/H_0$ is then imposed on all other assemblies. In addition, the condition of 10\% truncation of the smallest building block type in an assembly was desired to maintain flat top and bottom planes. 

Ideally, all tilings would be constructed to have the same number of block in each assembly.  However, the tiling structure imposes geometric constraints that such a condition cannot be met within a fixed $L_0/H_0$ value and the resulting bounded tilings range from 42 tiles up to 64 tiles, Table~\ref{tab:tilings} and  Fig.~\ref{fig:tilings}(c). Table~\ref{tab:tilings} lists all tile edge lengths values. The [3.4.6.4] tiling is the only tiling considered in this study possessing than one edge length value.

TIM systems require a bounding frame (fence) for constraint. The bounding frames were constructed by expanding each tiling beyond the boundaries drawn in Fig.~\ref{fig:tilings}(c) such that there exists a tile adjacent to every side of the outer tiles in the bounded set. Blocks were generated on these additional tiles such that the blocks formed from the bounded tiling were completely surrounded by this additional set of blocks. The blocks in the outermost set were fused into a single part to serve as a frame for the assembly.
The outer profile of this conglomerate frame was cut into either a square or hexagon shape as appropriate. 

The geometry of the single-tile systems is such that they can be flipped over and rotated to exactly overlay their original position. However, the multi-tile systems do not typically share this property.
The TIM system configurations used in this study are named after the bounded tiling from which they were created, and if the response of the assembly is direction dependent, the load direction will be indicated.
For example, the $[4^4]$-A assembly is not direction dependent, but the $[4.8^2]$-A assembly is, therefore it will be denoted as two separate configurations $[4.8^2]$-A(-) and $[4.8^2]$-A(+).
The complete set of TIM system configurations in this study is shown in Appendix A. 

\subsection{Analysis}
Finite element models are created for the analysis of the transverse force-deflection response. The bounding frame is considered as rigid and is fixed in space. Displacement boundary conditions are imposed via a rigid indenter pin interacting via contact located centrally to the assembly. A monotonically increasing displacement is applied to the indenter.  Individual building blocks are linear elastic and interact with each other by contact and friction. Details of the analysis approach are provided in Appendix B.
Calculations provide the force ($F$) -deflection ($u$) response of assemblies computed as the respective data on the indenter. 
The $F$-$u$ response is depicted as both raw and filtered data. System characteristic points are marked on the $F$-$u$ plots and these were extracted from each simulated configuration:
\begin{enumerate}
\item Stiffness as the secant to 80\% of the maximum force,
\item Strength as the maximum force recorded,
\item Displacement $u_{50}$ at the point the force drops to 50\% of its maximum value,
\item Toughness as the integral of $F$-$u$ up $u_{50}$, and
\item  Displacement $u_{slip}$ at the point the magnitude of the frictional dissipation becomes greater than the strain energy (\texttt{ALLFD} $>$ \texttt{ALLSE}).
\end{enumerate}

In all computations the mechanical response is dominated by the strain energy and friction dissipation such that the external work \texttt{ALLWK} is the sum of strain energy \texttt{ALLSE} and friction dissipation \texttt{ALLFD}. All other contributions (penalty work in the contact \texttt{ALLPW}, viscous dissipation \texttt{ALLVD}, artificial energies \texttt{ALLAE} and kinetic energy \texttt{ALLKE}) are negligible, at least up to conditions where slip starts to dominate and \texttt{ALLSE} $<$ \texttt{ALLFD}. 


\section{Results}\label{Results}
\begin{figure}[ht]
    \centering
    \subfigure[]{
        \includegraphics[width=0.35\textwidth]{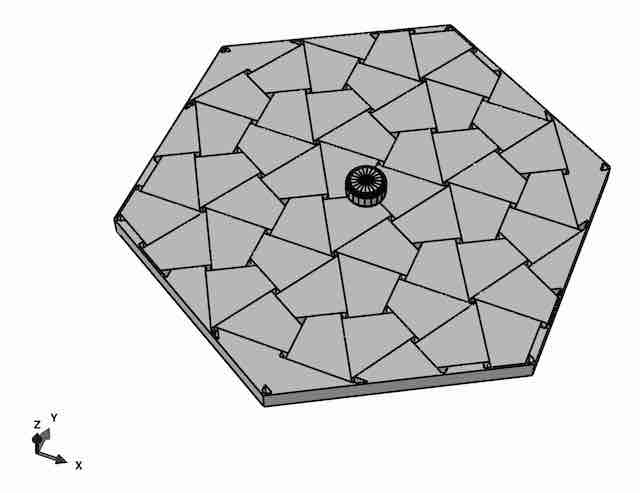}
         \label{3.4.6.4-B_config}
      }
      \subfigure[]{
        \includegraphics[width=0.35\textwidth]{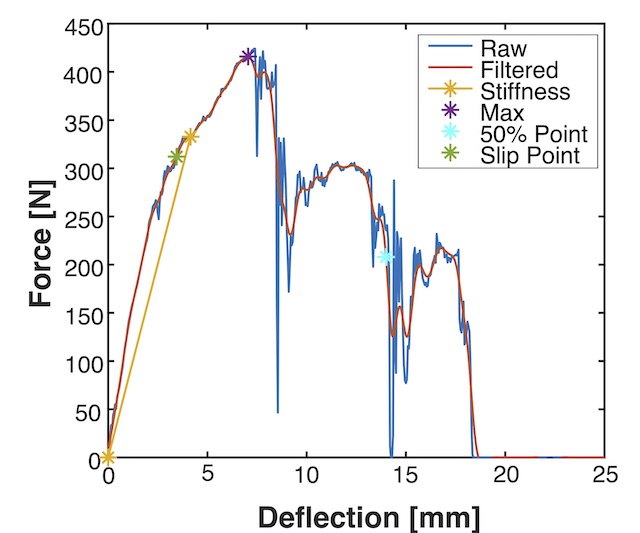}
        \label{3.4.6.4-B_FU}
        }
    \subfigure[]{
        \includegraphics[width=0.35\textwidth]{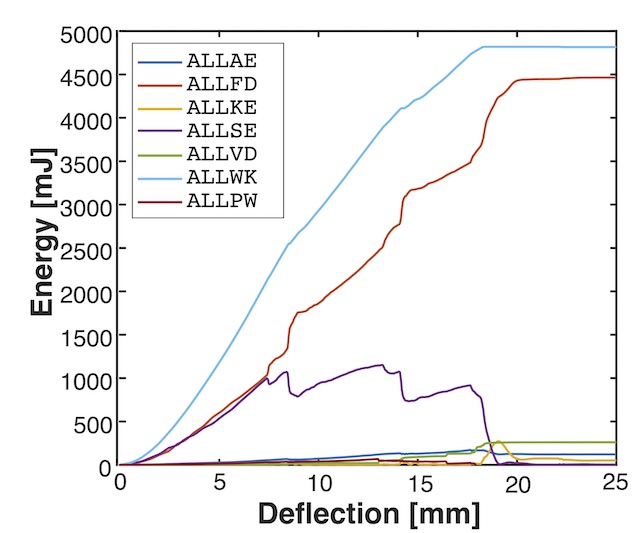}
        \label{3.4.6.4-B_energy}
        }
            \subfigure[]{
    \includegraphics[width=0.4\textwidth]{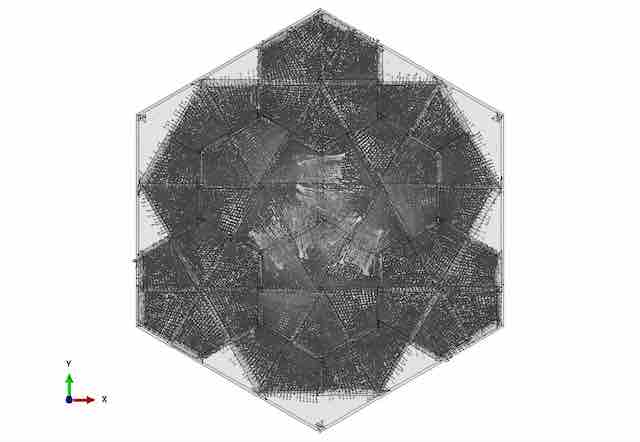}
    \label{3.4.6.4-B_vector}
    }
    \caption{(a) The $[3.4.6.4]$-B TIM system constructed with one tile type, (b) System energies, (c) Force-deflection response, (d) Vector plot of compressive principal stresses $\sigma_{p3}$ at the maximum load $\sigma_{p3}=[-28,+1]$MPa.}
    \label{onetile-results}
\end{figure}

\begin{figure}[ht!]
    \centering
    \subfigure[]{
      \includegraphics[width=0.35\textwidth]{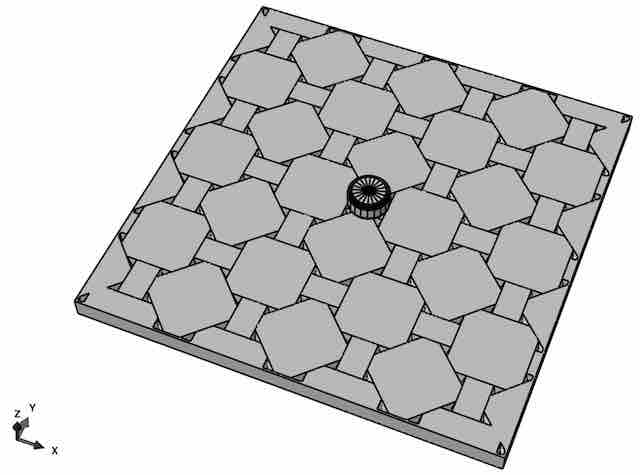}
    \label{4.8^2-B_config}
      }
      \subfigure[]{
        \includegraphics[width=0.35\textwidth]{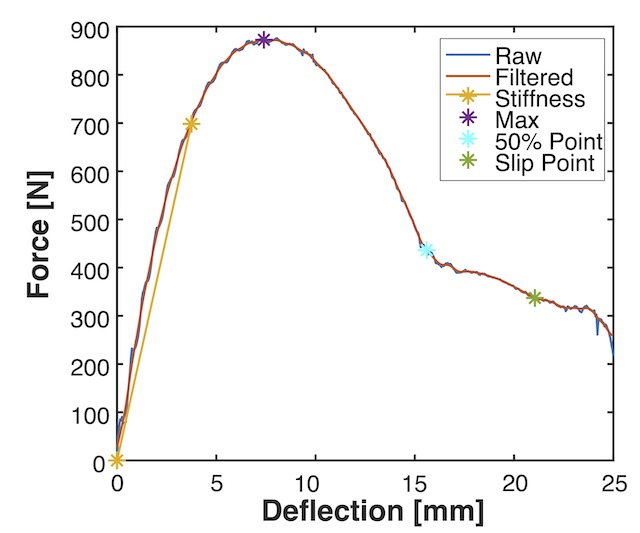}
        \label{4.8^2-B_FU}
        }
    \subfigure[]{
        \includegraphics[width=0.35\textwidth]{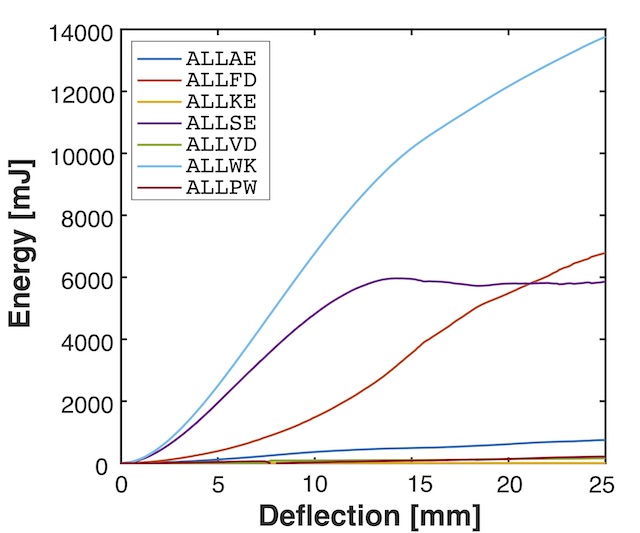}
        \label{4.8^2-B_energy}
        }
            \subfigure[]{
     \includegraphics[width=0.4\textwidth]{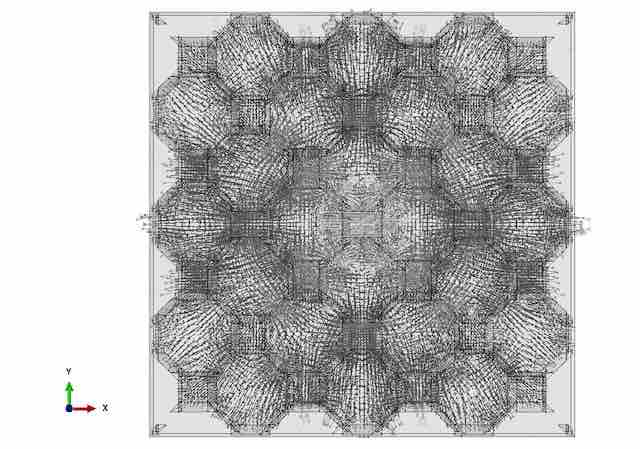}
    \label{4.8^2-B_vector}
    }
    \caption{(a) The $[4.8^2]$-B TIM system constructed with two tile types, (b) System energies (c) Force-deflection response, Vector plot of compressive principal stresses $\sigma_{p3}$ at the maximum load with $\sigma_{p3}=[-52,+2]$MPa.}
    \label{twotiles-results}
\end{figure}

\begin{figure}[ht!]
    \centering
    \subfigure[]{
    \includegraphics[width=0.35\textwidth]{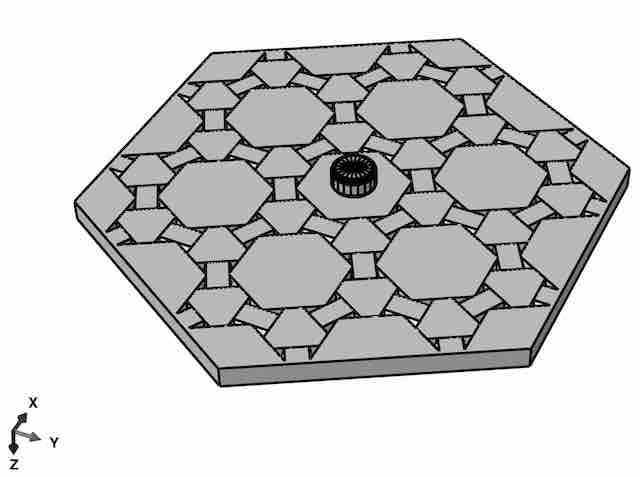}
    \label{4.6.12.A+_config}
      }
      \subfigure[]{
        \includegraphics[width=0.35\textwidth]{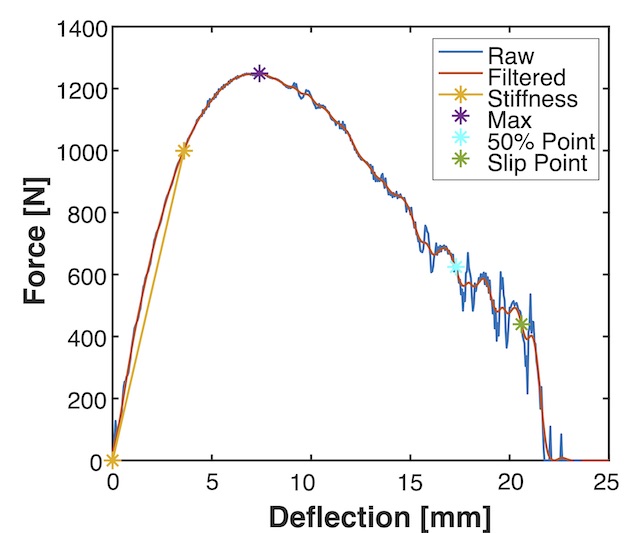}
        \label{4.6.12.A+_FU}
        }
    \subfigure[]{
        \includegraphics[width=0.35\textwidth]{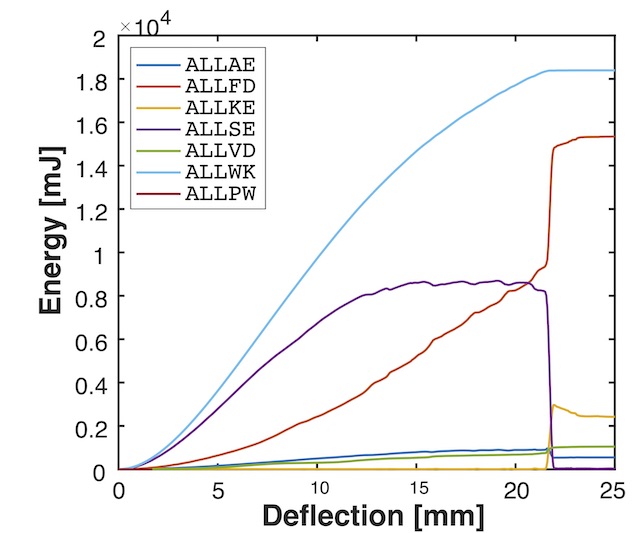}
        \label{4.6.12.A+_energy}
        }
            \subfigure[]{
     \includegraphics[width=0.4\textwidth]{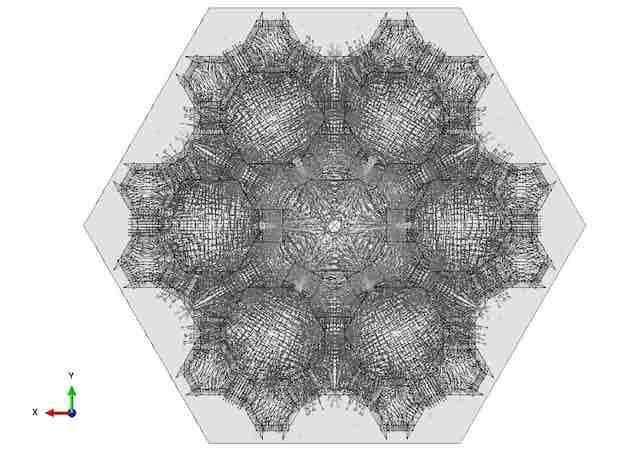}
    \label{4.6.12.A+_vector}
    }
    \caption{(a) The $[4.6.12]$-A(+) TIM system constructed with three tile types, (b) System energies, (c) Force-deflection response, (d) Vector plot of compressive principal stresses $\sigma_{p3}$ at the maximum load with $\sigma_{p3}=[-120,+1]$MPa.}
    \label{threetiles-results}
\end{figure}

Results for computations for (\emph{i}) the $[3.4.6.4]$-B system with a single tile type,  Fig.~\ref{3.4.6.4-B_config}, (\emph{ii}) the $[4.8^2]$-A(+) system with two different tiles,  Fig.~\ref{4.8^2-B_config} , and (\emph{iii}) the $[4.6.12]$-A(+) system composed of three different tiles, Fig.~\ref{4.6.12.A+_config}, are depicted as representative for the computational study.

 The $F$-$u$ curves (Figs~\ref{3.4.6.4-B_FU},~\ref{4.8^2-B_FU},~\ref{4.6.12.A+_FU}) overall possess the skewed parabola shape with a gradual load decrease past the maximum load, similar to what has been documented in other investigations on TIM systems \cite{mather2012structural,Valashani2015ANacre,Dyskin2003FractureContacts,Mirkhalaf2018SimultaneousCeramics}. Stiffness, strength, toughness, the rate of force drop post the load maximum, and the slip onset vary distinctly between assembly architectures.
 Initially, the $F$-$u$ curves are smooth and deformation is by tilting of unit blocks and their elastic deformation. As deformation progresses, local slip events become apparent in the $F$-$u$ curve as intermittent load drops. 
 
 The three examples depicted represent conditions where the onset of slip dominance is significantly different. The contribution of slip to the deformation response can be assessed from the evolution of the systems energies during loading, Figs~\ref{3.4.6.4-B_energy},~\ref{4.8^2-B_energy},~\ref{4.6.12.A+_energy}. For the $[3.4.6.4]$-B case, slip is a strongly dominant factor. Friction dissipation is of equal magnitude as the strain energy already during early stages of loading, and slip becomes dominant past the maximum load at $u=7.58$ mm. 
 For the $[4.8^2]$-B configuration and the $[4.6.12]$-A(+) case, the strain energy is much larger than the frictional dissipation over much of the load history. The slip onset condition is delayed to $u_{slip}$=21.05 mm for the $[4.8^2]$-B case and to $u_{slip}$=20.6 mm $[4.6.12]$-A(+) case, far into the deformation histories. Slip alone is not the sole determining factor for the strength of a system. While the $[3.4.6.4]$-B with the largest slip contribution also possesses the lowest strength $F_{max}=$415.9 N, the two other systems possess distinctly different strength despite similarly delayed slip: $F_{max}$ for $[4.8^2]$-B is 872.7 N and for $[4.6.12]$-A(+) it is 1249.0 N. Past the maximum load, failure is gradual at least until the latest stages of the deformation history. For the $[3.4.6.4]$-B assembly, strong local intermittent load drops are associate with slip events and load carrying capacity is lost early, ($u_{50}=$14.0 mm). For the other two assemblies slip events are less pronounced in the $F$-$u$ data, and $u_{50}$ values are significantly larger: $u_{50}$=15.6 mm   for $[4.8^2]$-B and 17.3 mm for $[4.6.12]$-A(+). As a consequence, toughness values are also significantly different. The toughness is the least for $[3.4.6.4]$-B, followed by $[4.8^2]$-B and $[4.6.12]$-A(+). 
 
 In TIM systems, load transfer is dominated by compressive loads in building blocks balanced by tensile loads in the bounding frame.
 The computed load transfer patterns in the assemblies are depicted as vector plots of the compressive principal stress $\sigma_{p3}$ at the state of maximum load.
 In the assembly $[3.4.6.4]$-B the distribution of $\sigma_{p3}$ is found to be rather homogeneous throughout and the entire assembly perimeter transfers load to the bounding frame, Fig.~\ref{3.4.6.4-B_vector}.
 For the assembly $[4.8^2]$-B it is found that $\sigma_{p3}$ is less in the larger tiles than it is in the smaller ones and a distinct load transfer pattern is seen, Fig.~\ref{4.8^2-B_vector}.
 Now loads are transferred to the frame only along a subset of faces to the bounding frame but both types of tiles contribute.
 The finding  of load transfer being dominant via the smallest building blocks is also present in the results for the $[4.6.12]$-A(+) assembly, Fig.~\ref{4.6.12.A+_vector}.
 In this case load transfer to the frame is found to occur predominantly via the faces of the smallest building blocks.
 
\begin{figure}[ht]
    \centering
    \subfigure[]{\includegraphics[width=0.45\textwidth]{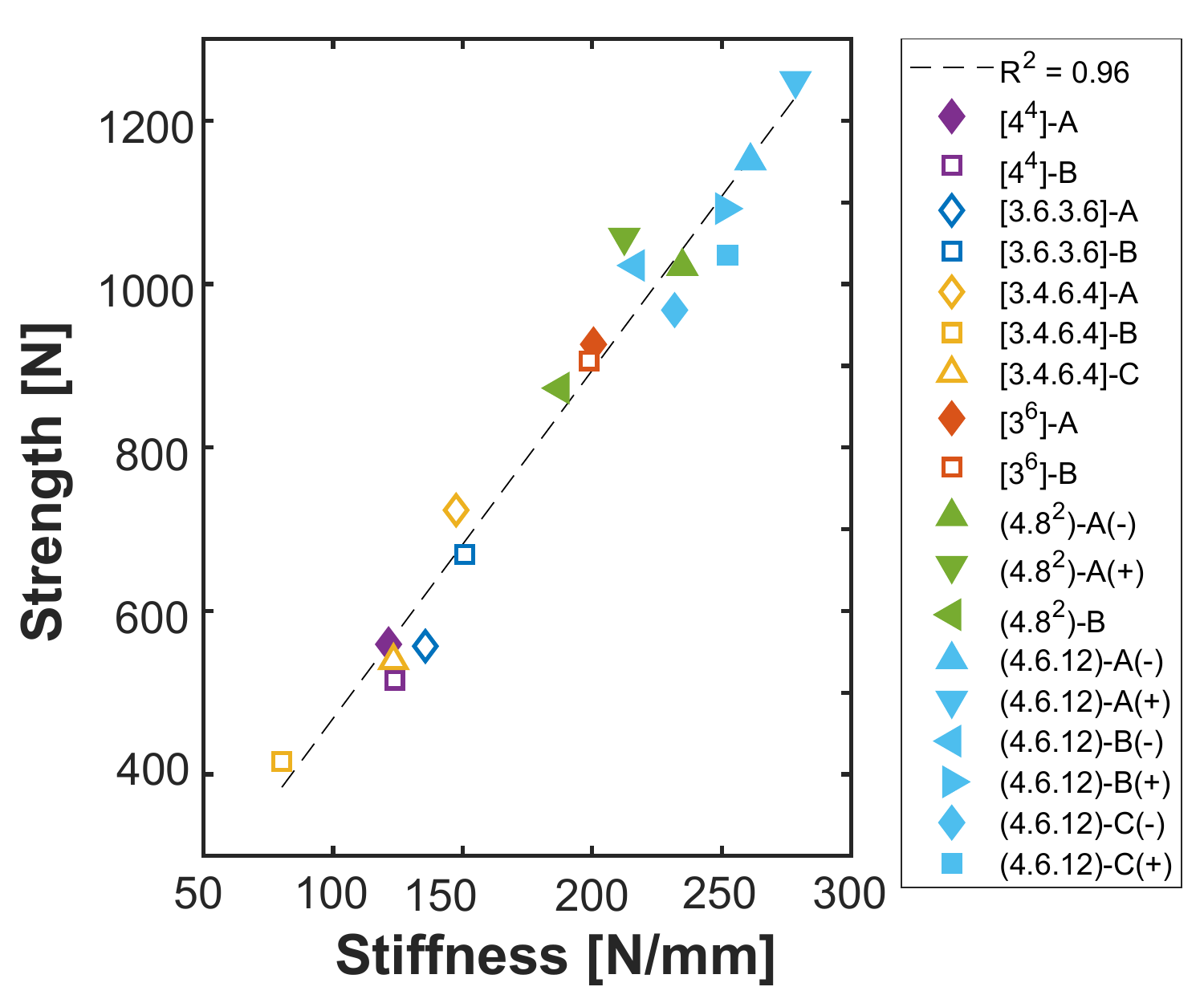}
     \label{corr_SK}
    }
     \subfigure[]{\includegraphics[width=0.45\textwidth]{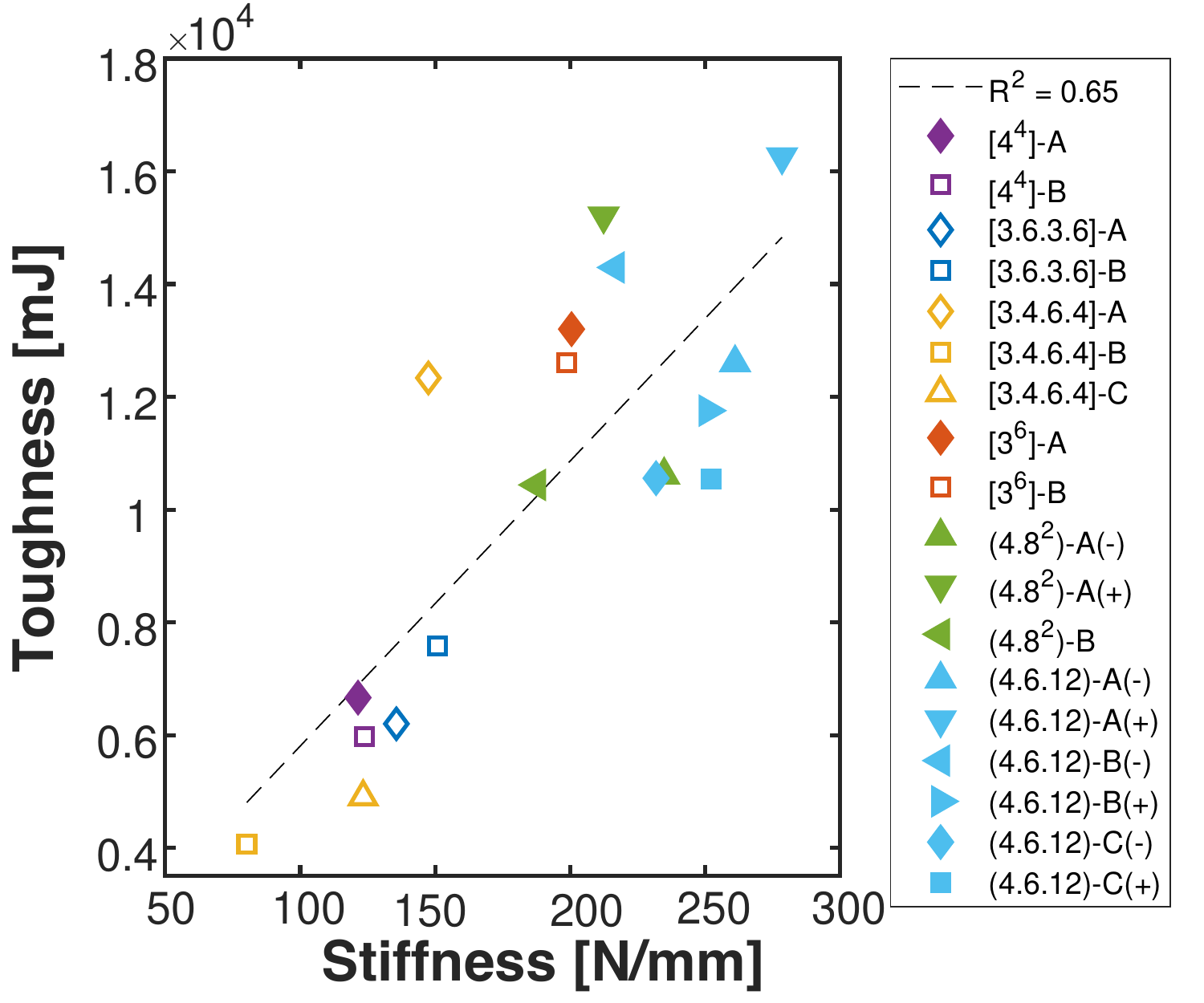}
      \label{corr_TK}
    }
    \subfigure[]{\includegraphics[width=0.45\textwidth]{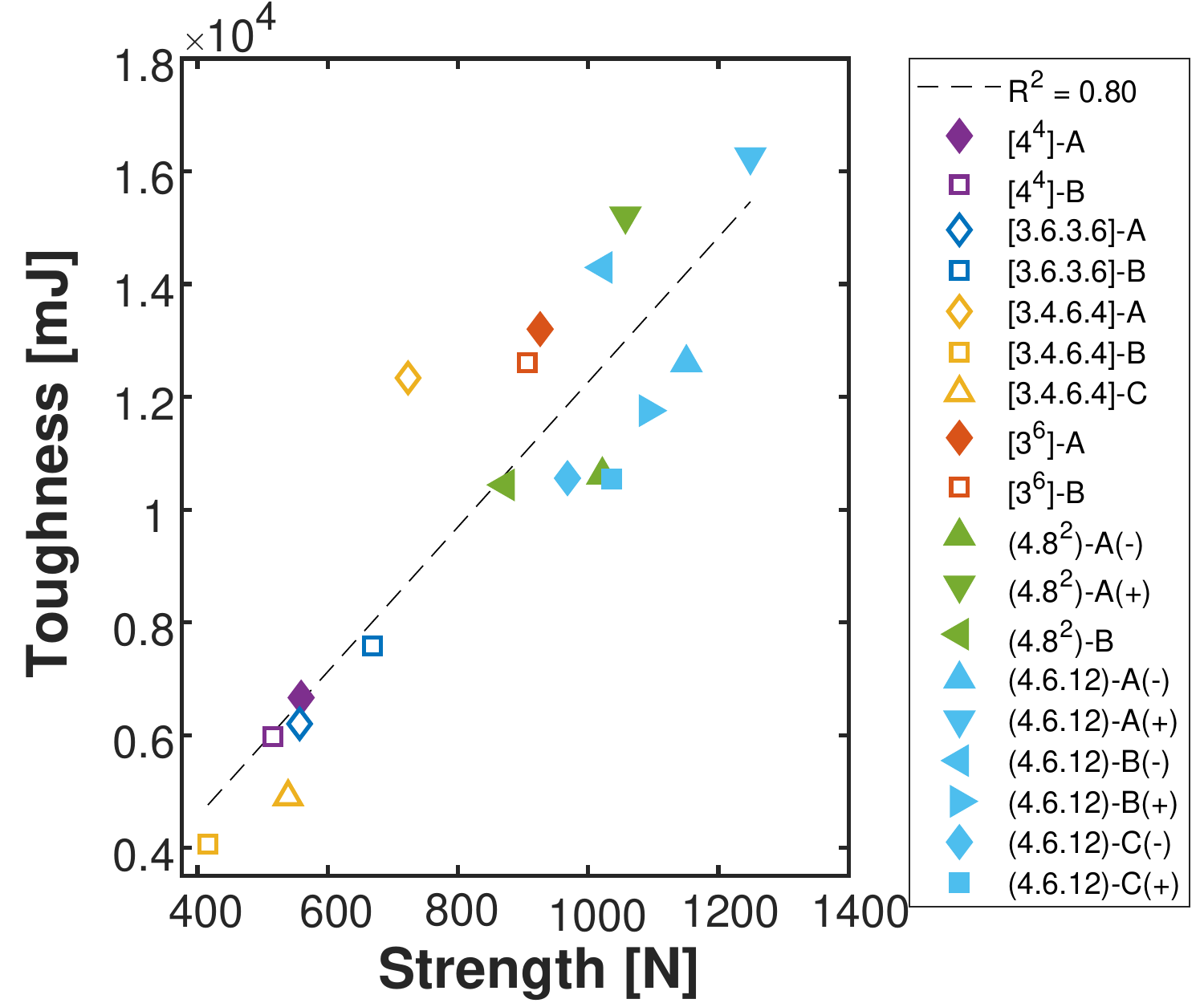}
     \label{corr_TS}
    }
\subfigure[]{\includegraphics[width=0.45\textwidth]{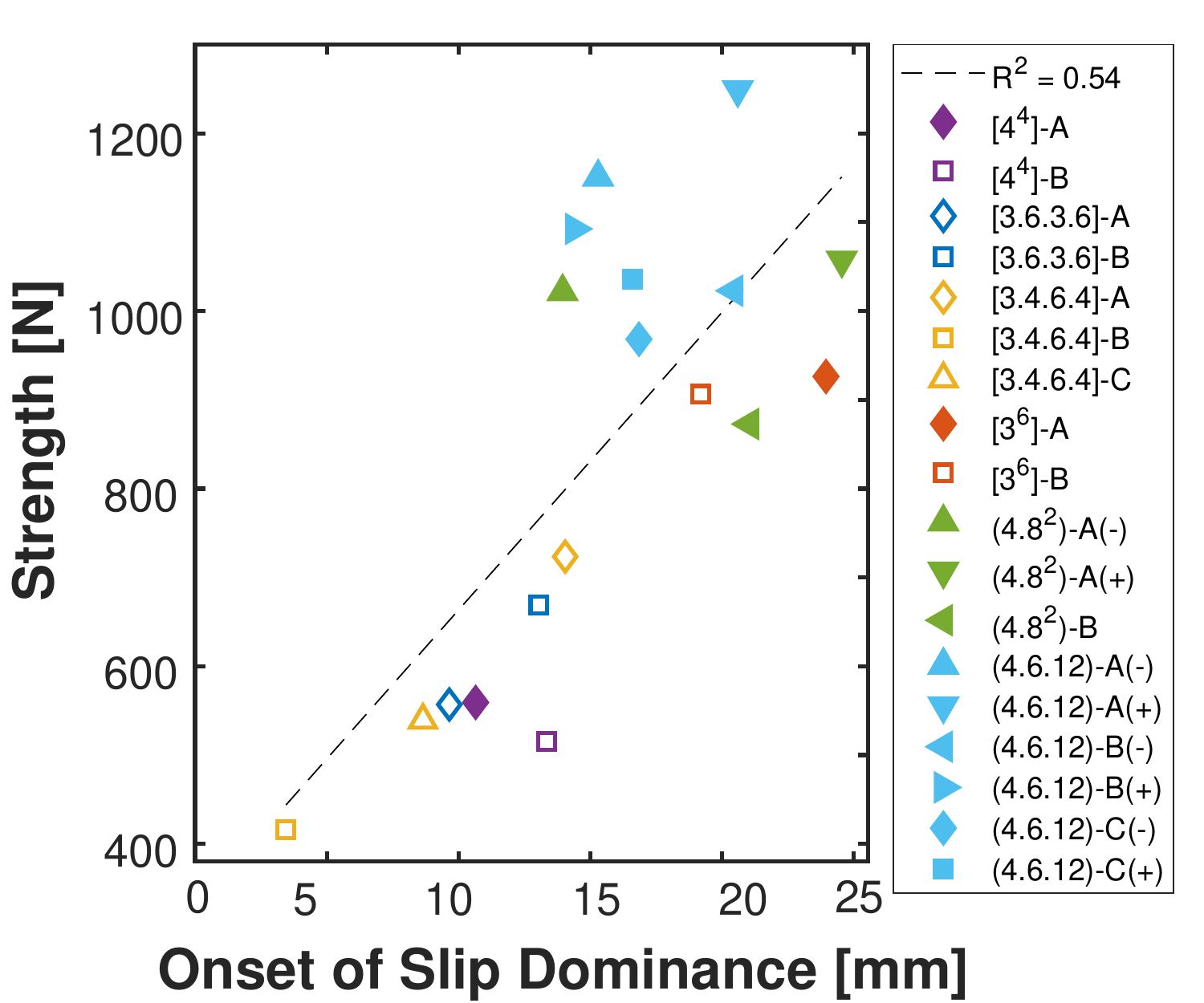}
 \label{corr_K-slipOnset}
    }
     \caption{Cross property relationships: (a) Strength and stiffness, (b) Toughness and stiffness (c) Toughness and strength (d) Stiffness and onset of slip dominance.}
\end{figure}

Subsequently, the characteristics of all computed configurations are considered in the form of cross-property relationships. Strength and stiffness were linearly correlated to a high degree, Fig.~\ref{corr_SK}. 
Stiffness and toughness, Fig.~\ref{corr_TK}, ($R^2$=0.65) as well as strength and toughness, Fig.~\ref{corr_TS}, ($R^2$=0.80) are also linearly correlated, but at a somewhat smaller $R^2$ value.  From the results Figs~\ref{onetile-results}, ~\ref{twotiles-results} and ~\ref{threetiles-results} it could be inferred that the prevalence of slip would be a good predictor of TIM properties. However, this was found to be only partially the case. Strength is related to $u_{slip}$ but the correlation is weak, Fig.~\ref{corr_K-slipOnset} ($R^2$=0.64). The relationship between stiffness and $u_{slip}$ is even weaker at $R^2=0.43$.

\section{Discussion}
The TIM plate systems introduced in this study exhibit attractive properties in terms of their failure. A gradual decrease in load is realized even if the material used to make the building blocks would be considered as brittle itself. Such a response is found across all system architectures considered.

The overall transverse force-deflection response of TIM systems has been explained by the formation of multiple force chains in the granular-like assembly.
Such force chains reach from the top plate face at which the load is applied to the opposing bottom plate face.
As the plate deflection increases so do the angles between the force chains and the plate reference plane. 
Such a process is similar to what happens in a Mises-truss.
In \cite{short2019scaling} a comprehensive model for this approach is demonstrated.
Thus, the Mises-truss model is used to describe the force-deflection response of TIM plates, rather than a monolithic plate theory. 
Within the Mises-truss model, the observation of a linear dependence of strength on stiffness is consistent.
The finding of rather linear relationships between stiffness and toughness and also between strength and toughness again relates well to the Mises-truss model, as does the observation that the displacement to final failure is rather constant and in the range of 2.0 - 2.5 times the assembly thickness.
The variations in strengths (from about 400 to 1250 N) result in a vertical stretching of the skewed parabola $F$-$u$ response in which the stiffness and strength vary while the deflection to failure does not.

It is of interest to relate the system mechanical characteristics to the architectural aspects in order to find predictors of system performance.
As strength is well correlated with both stiffness and toughness, a predictor for strength is also capable of predicting stiffness and toughness. To start, strength was correlated against the number of tiles in an assembly, Table 1. It was desired that there not be a correlation between these two parameters. With a coefficient of determination $R^2 = 0.02$, this goal was met.
Next, strength was correlated against segmentation in the assembly.

\begin{figure}[ht]
    \centering
    \subfigure[]{\includegraphics[width=0.45\textwidth]{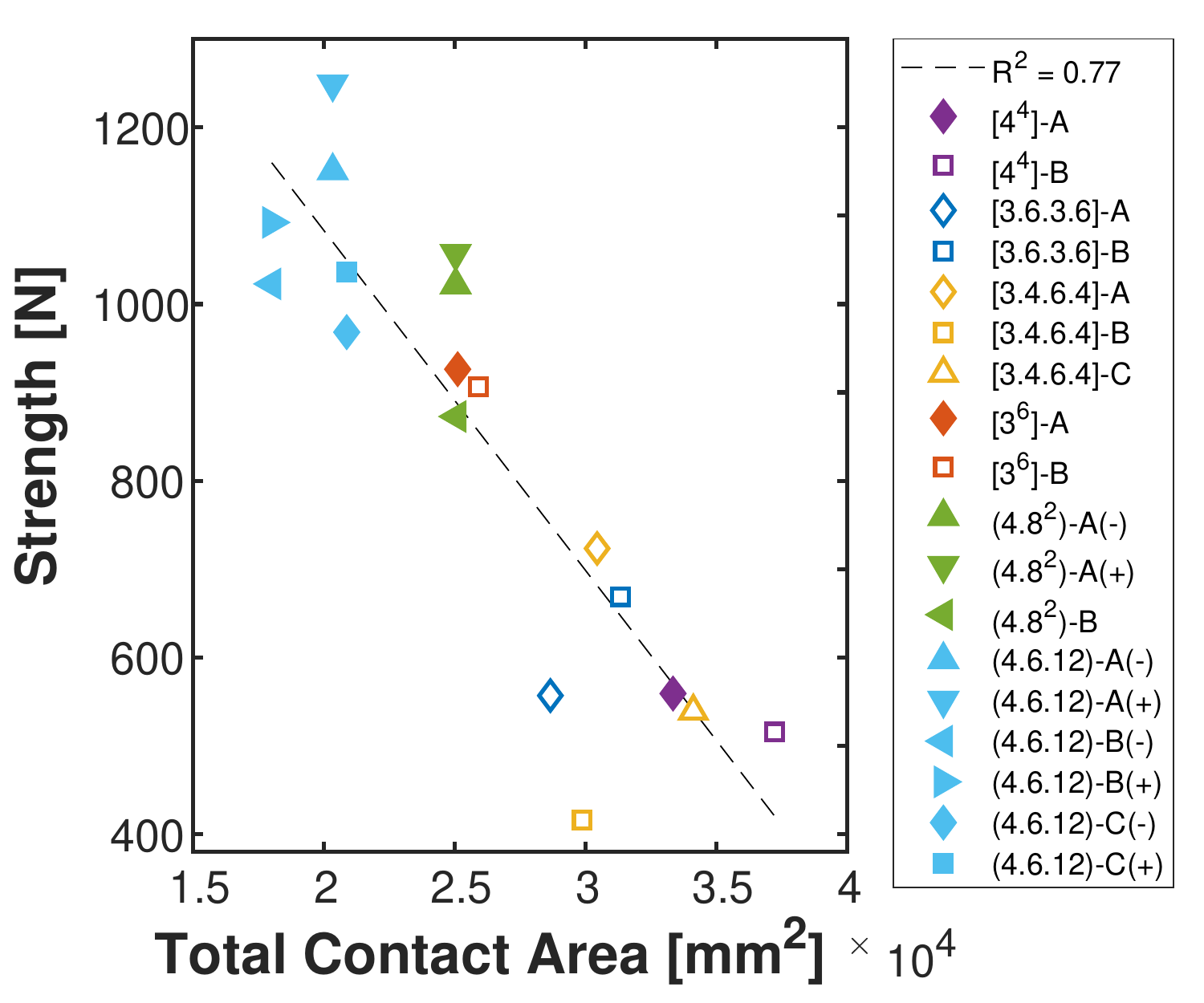}
    \label{corr_SATotal}
    }
    \subfigure[]{\includegraphics[width=0.45\textwidth]{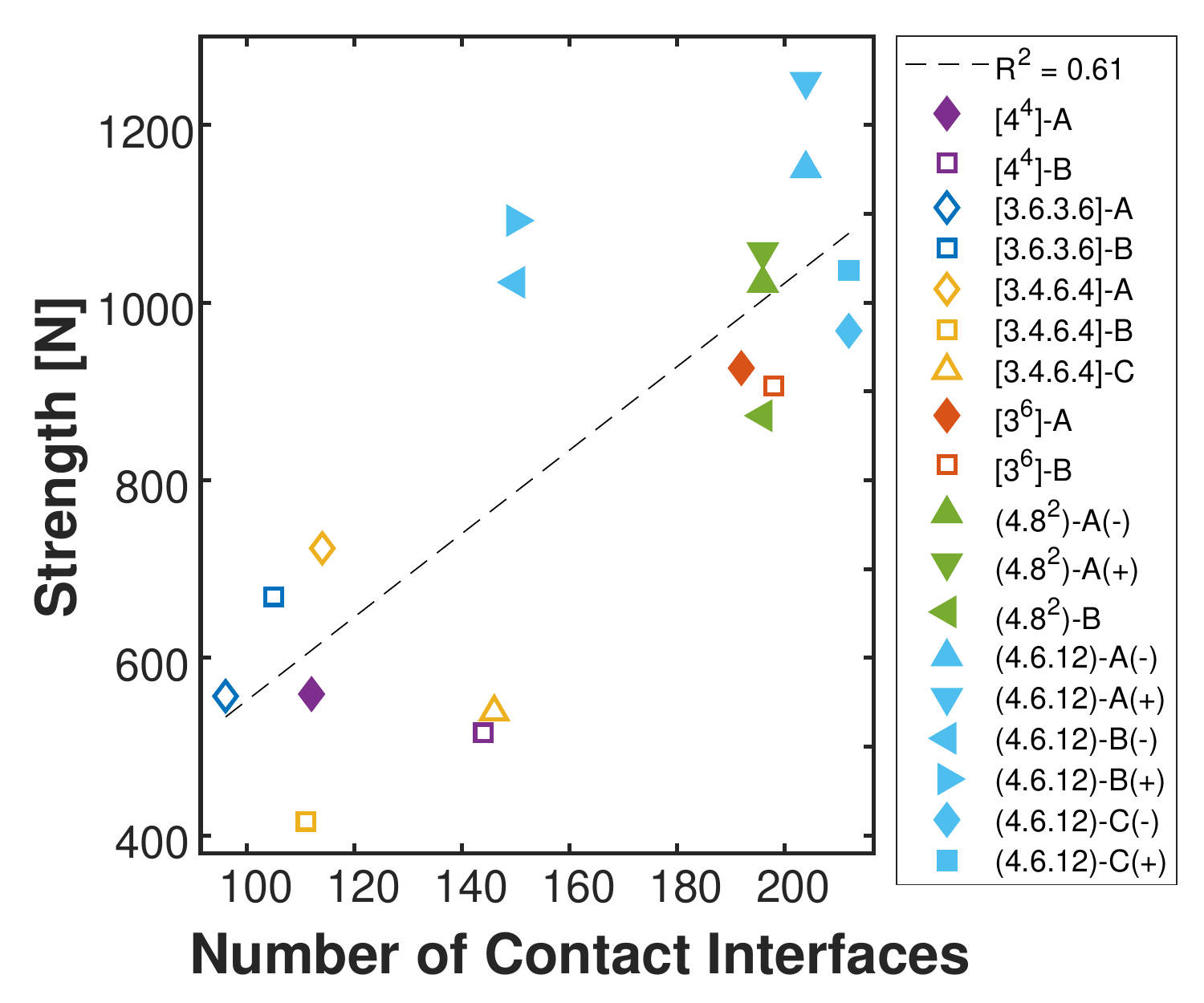}
    \label{corr_SNumContact}
    }
    \subfigure[]{\includegraphics[width=0.45\textwidth]{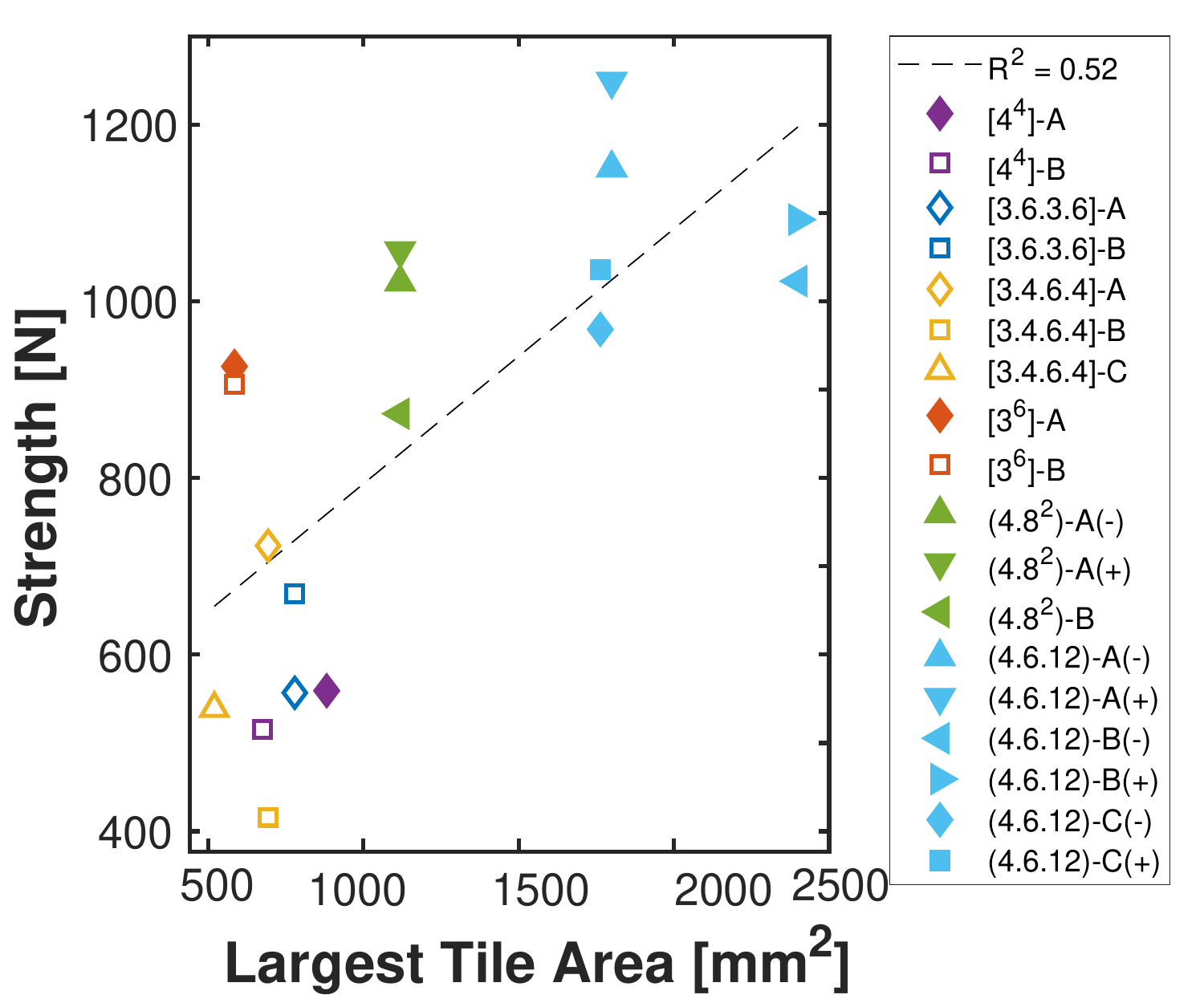}
    \label{corr_SAMax}
    }
    \subfigure[]{\includegraphics[width=0.45\textwidth]{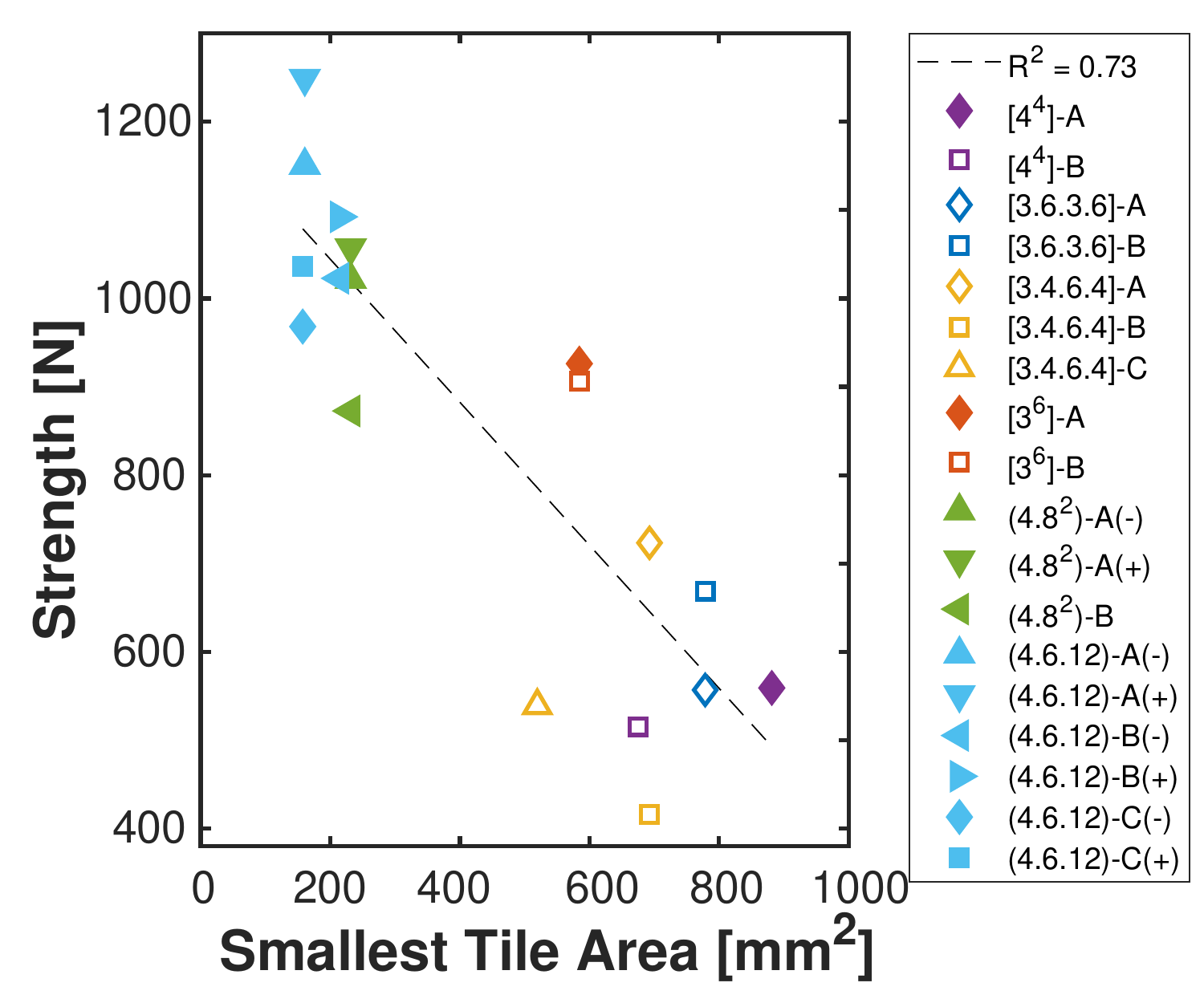}
    \label{corr_SAMin}
    }
    \caption{(a) Strength vs total contact area between segmented bodies; (b)Strength vs number of contact interfaces between segmented bodies; (c) Strength vs area of the largest tile in the base tiling; (d) Strength vs area of the smallest tile in the base tiling.}
      \label{correlations}
    \end{figure}

One measure of the degree of segmentation is the total contact area between all segmented bodies in the assembly. The total contact area between all segmented bodies in the assembly is computed in the assembly's initial position before any displacement had occurred.
Strength and total contact area between all segmented bodies in the assembly are well correlated, Fig.~\ref{corr_SATotal} ($R^2=0.77$).  Smaller values of total contact area lead to higher strength. This suggests that the less segmented a structure is, the greater its strength will be. This argument would intuitively agree with the fact that a monolithic plate is generally stronger than its segmented counterparts.

A second measure of the degree of segmentation is the number of contact interfaces in the assembly, defined as a state of contact between any two bodies.
Strength and the number of contact interfaces are less significantly related, Fig.~\ref{corr_SNumContact} ($R^2=0.61$), but an increase in the number of contact interfaces is correlated to an increase in strength suggesting that the more segmented a structure is, the greater its strength will be.

The correlations between strength and the two measures of segmentation are in obvious disagreement.
Clearly the degree of segmentation alone is insufficient in predicting the properties of the TIM systems under consideration here.
The present data suggests that TIM behavior must be dependent on how the system is segmented rather than how much it is segmented.

The assemblies having a larger number of contact interfaces did so by having building blocks with a greater number of sides.
It is possible to increase the number of contact interfaces by increasing the number of building blocks, but the TIM systems in this study all had approximately the same number of building blocks.
Therefore, the increase in strength seen with the increase in contact interfaces might be attributed to the presence of larger building blocks, rather than to the increase in contact interfaces.
This might describe the gap in strength values that is seen between the weakest configurations (all based on tessellations with a single four-sided tile) and the other configurations which also include larger tiles with a larger number of sides, Fig.~\ref{corr_SNumContact}.

To further investigate the dependence of the mechanical behavior on system architecture, strength is correlated against the area of the largest tile in the tessellation from which each TIM system was constructed, Fig.~\ref{corr_SAMax}. Strength is found as positively correlated with the largest tile area in the assembly ($R^2=0.52$). This finding does support the previous conjecture that TIM systems with larger blocks would be stronger, but with a coefficient of determination $R^2 = 0.52$ it is not a strong correlation.
Given that the bounded tilings in this study all have approximately the same number of tiles and about the same total area, if there are larger tiles in a tiling, it must also possess some smaller tiles. Thus, strength is correlated against the area of the smallest tile in the tessellation from which each TIM system was constructed, Fig.~\ref{corr_SAMin}.
This relationship possesses a coefficient of determination $R^2 = 0.73$ and suggests that the smallest tile size is a better predictor of strength than the largest tile size.

The findings on architecture-property relationships suggest that the strongest TIM systems are the ones with the least total contact area, the greatest number of contact interfaces, and the smallest tiles. This combination of characteristics leads to the conclusion that TIM system configurations having architectures that constrict load transfer into well defined force chains possess the greatest strength. These increasing degrees in the concentration of force chains are well represented in the results of the computations for the TIM configurations based on the $[3.4.6.4]$-B, $[4.8^2]$-B, and $[4.6.12]$-A tilings, Figs~\ref{3.4.6.4-B_vector},~\ref{4.8^2-B_vector},~\ref{4.6.12.A+_vector}. 
A more distinctly developed force chain network is thereby not related only to the presence of smaller tiles in the assembly but also to the geometry of the tessellation as the force chain network structure develops within a specific assembly.

The present results were confined by two constraints: assemblies are planar and the interlocking geometry is based on planar tile faces. Neither constraint is seen as a limitation in the application of present results. The geometric arguments on tessellations and the resulting assemblies overall can certainly be extended to curved systems made of topologically interlocked building blocks which have recently been demonstrated in the context of digital design and manufacturing approaches \cite{Bejarano2019AConfigurations,fallacara2006digital,block2017armadillo}. As it has been demonstrated that osteomorphic shaped interlocking \cite{molotnikov2007percolation,oikonomopoulou2019experimental} and multiscale interlocking  \cite{djumas2017deformation} provide similar or improved mechanical response as planar type interlocking, it can be argued that the present results will be applicable to such systems as well.

\section{Conclusion}\label{conclusions}

TIM systems were constructed for 18 configurations based on six unique tilings and their response under transversely applied displacement load is investigated.
It was found that the load responses of all configurations were generally consistent with the typical skewed parabola that has been recorded in other TIM systems.
The attractive positive correlations of toughness-stiffness and toughness-strength were realized for all configurations.
There exists significant variance in the performance of the TIM systems in this study.
It was generally observed that the triple-tile $(4.6.12)$ configurations were the strongest, followed but the double-tile $(4.8^2)$ configurations, the single hexagon tile $[3^6]$ configurations, and then all the single four-sided tile configurations.
The stiffest, strongest, and toughest configurations tended to have the least total contact area between segmented bodies, the greatest number of contact interfaces, and the smallest tiles.
It is postulated that this combination of features leads to more confined force chains of the internal load transfer.
The findings of this study allow for an expansion of the material space.
When considering a segmented material system, a greater range of ductility is available as compared to homogeneous materials.
The tessellation pattern can be chosen to achieve the desired ductility
These methods can be used to design advantageous material systems that are ductile as a system while maintaining high strength within the individual components.\\

\noindent \textbf{Acknowledgment:} This work supported by the National Science Foundation under Grant No. 1662177.

\newpage

\appendix
\section{TIM Assemblies}
Figures~\ref{44_assembly} to Fig.~\ref{4612_assembly} depict the TIM systems under consideration. 
\newcommand{\assemblyPlotA}[1]{
    \includegraphics[width=0.3\textwidth,trim={120 50 700 70},clip]{#1}}

\begin{figure}[ht]
    \centering
    \subfigure[]{
        \assemblyPlotA{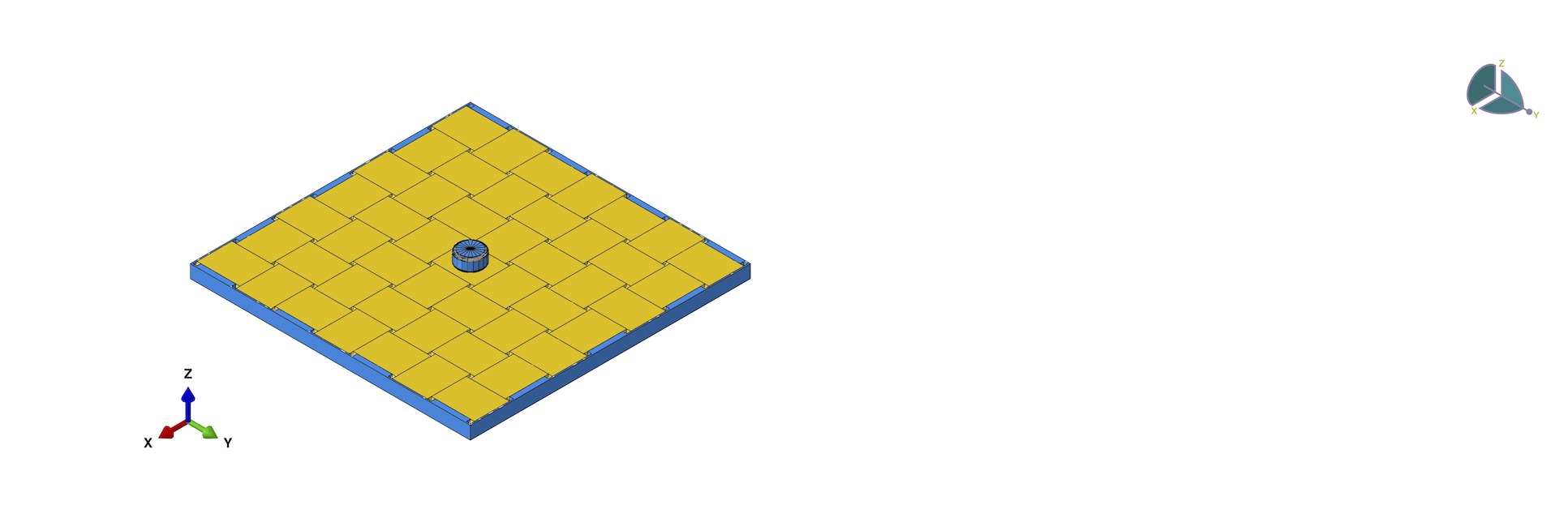}
        \label{44-A-1_assembly}}
    \subfigure[]{
        \assemblyPlotA{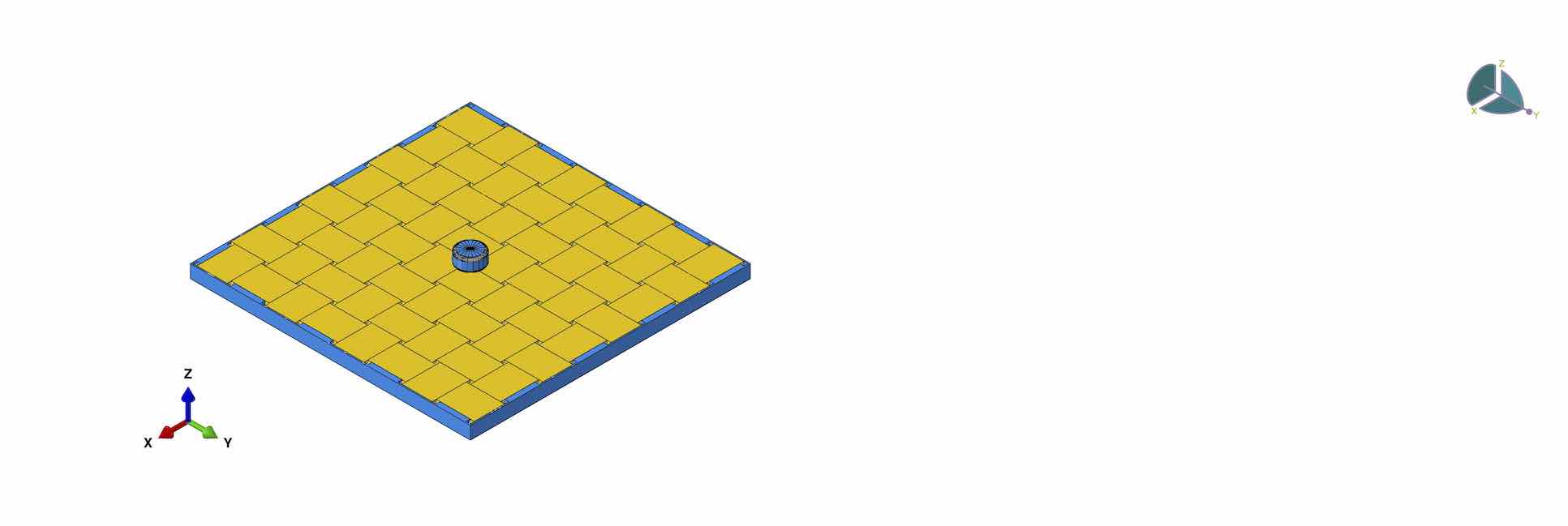}
        \label{44-B-1_assembly}}
    \caption{Assembly configurations: (a) $[4^4]$-A, (b) $[4^4]$-B.}
    \label{44_assembly}
\end{figure}

\begin{figure}[ht]
    \centering
    \subfigure[]{
        \assemblyPlotA{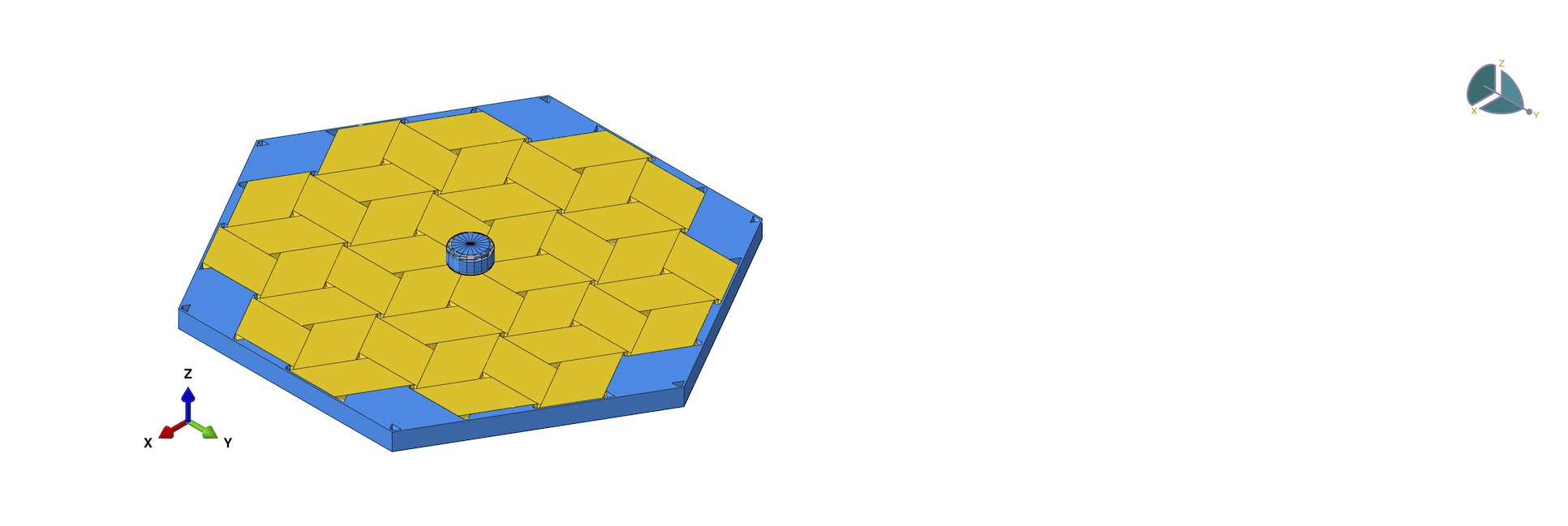}
        \label{3636-A-1_assembly}}
    \subfigure[]{
        \assemblyPlotA{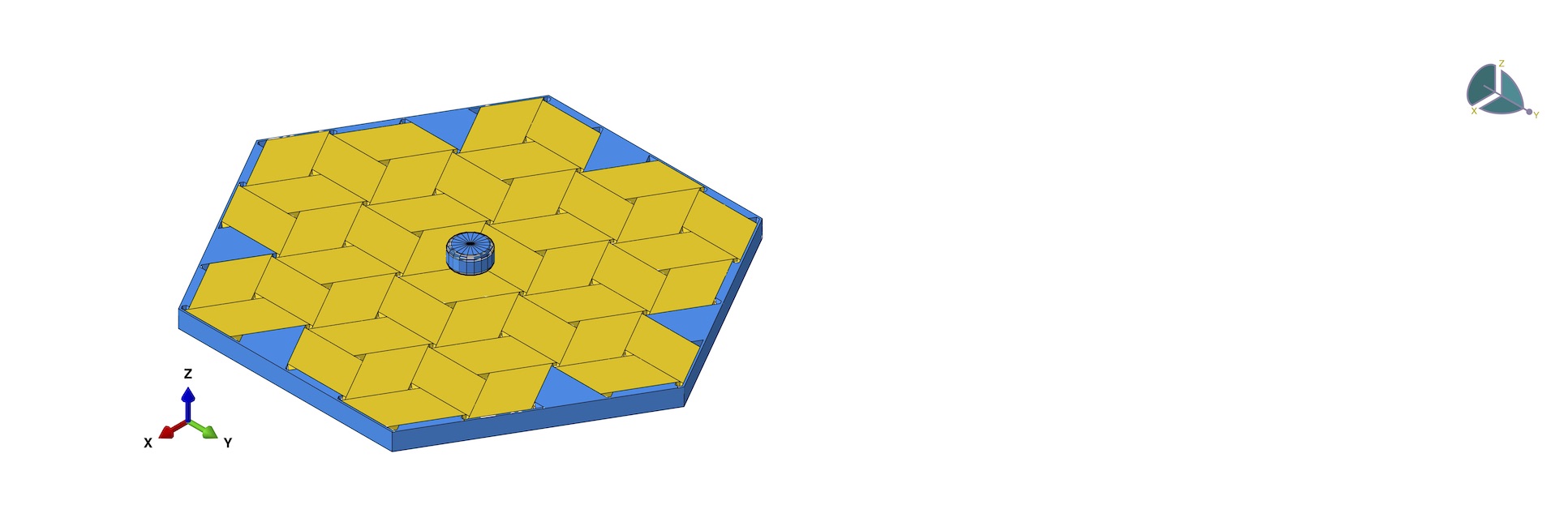}
        \label{3636-B-1_assembly}}
    \caption{Assembly configurations: (a) [3.6.3.6]-A, (b) [3.6.3.6]-B.}
    \label{3636_assembly}
\end{figure}

\begin{figure}[ht]
    \centering
    \subfigure[]{
        \assemblyPlotA{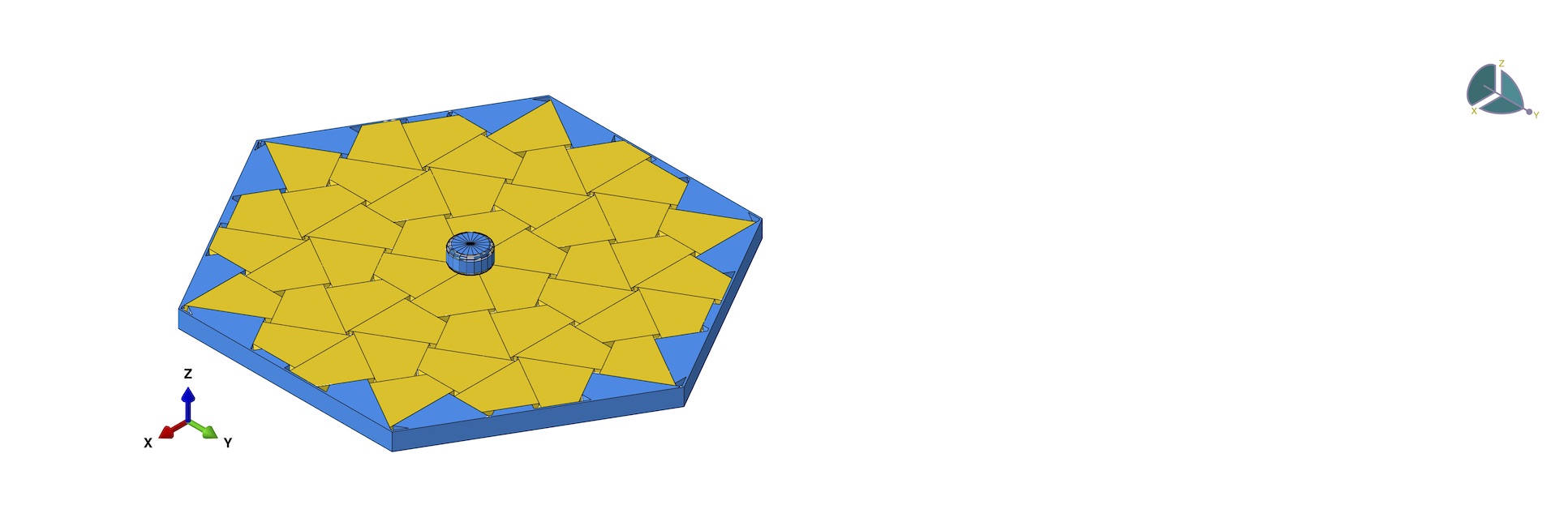}
        \label{3464-A-1_assembly}}
    \subfigure[]{
        \assemblyPlotA{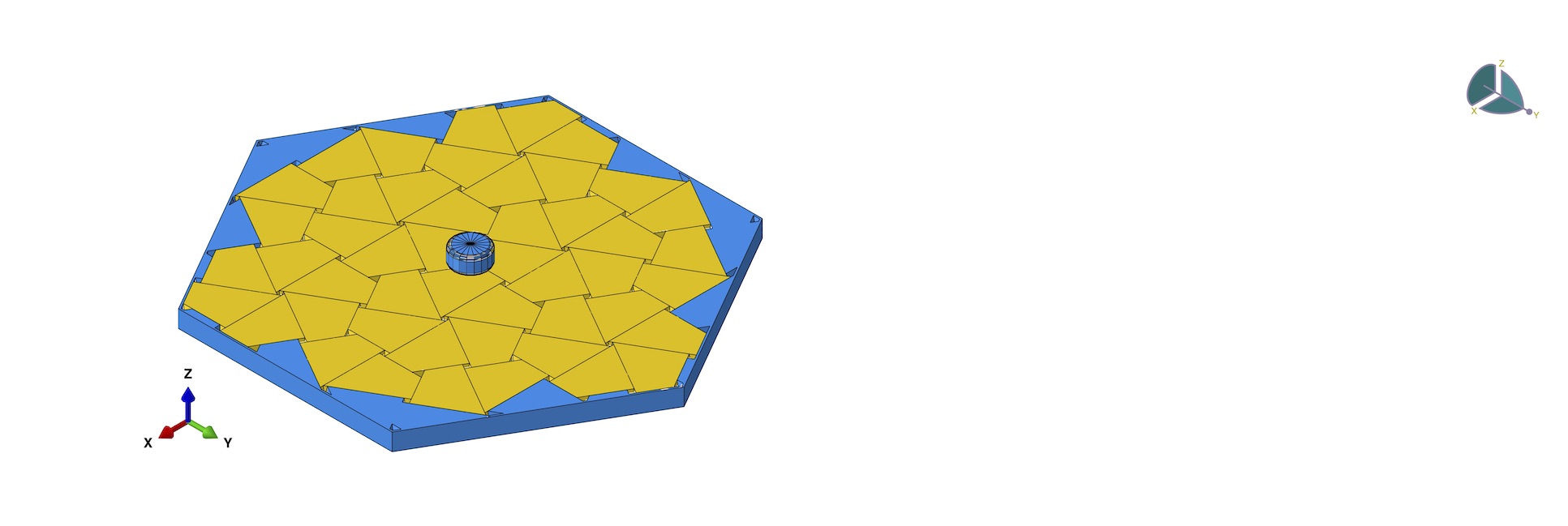}
        \label{3464-B-1_assembly}}
    \subfigure[]{
        \assemblyPlotA{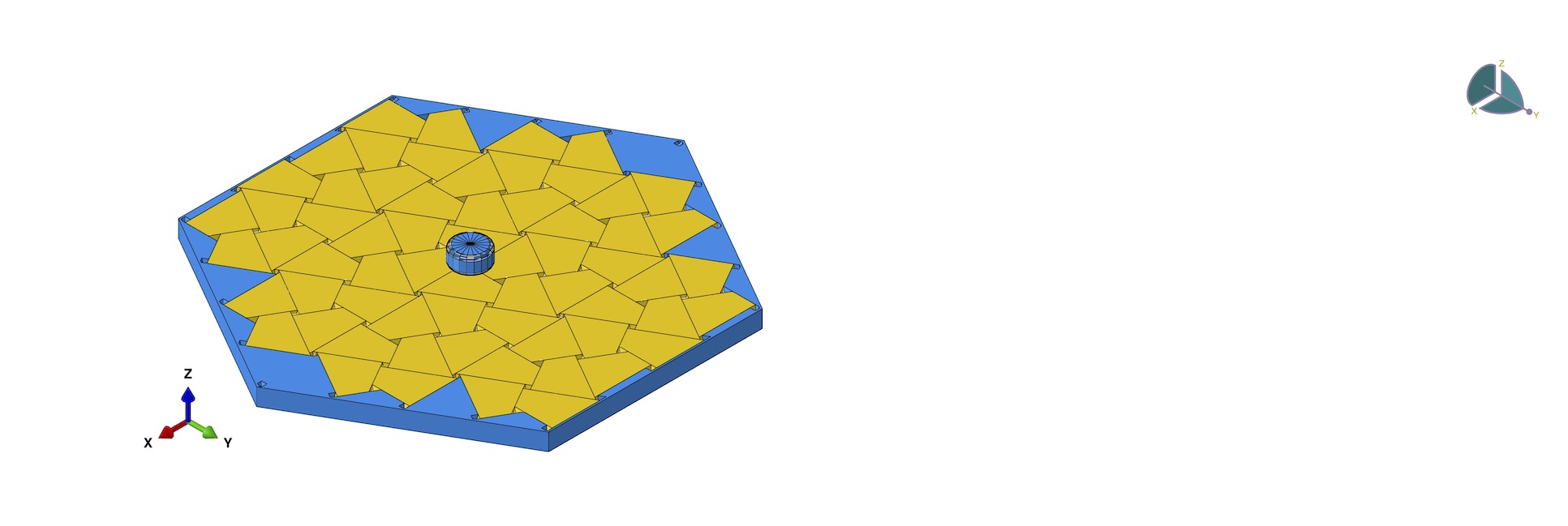}
        \label{3464-C-1_assembly}}
    \caption{Assembly configurations: (a) [3.4.6.4]-A, (b) [3.4.6.4]-B, (c) [3.4.6.4]-C.}
    \label{36464_assembly}
\end{figure}

\begin{figure}[ht]
    \centering
    \subfigure[]{
        \assemblyPlotA{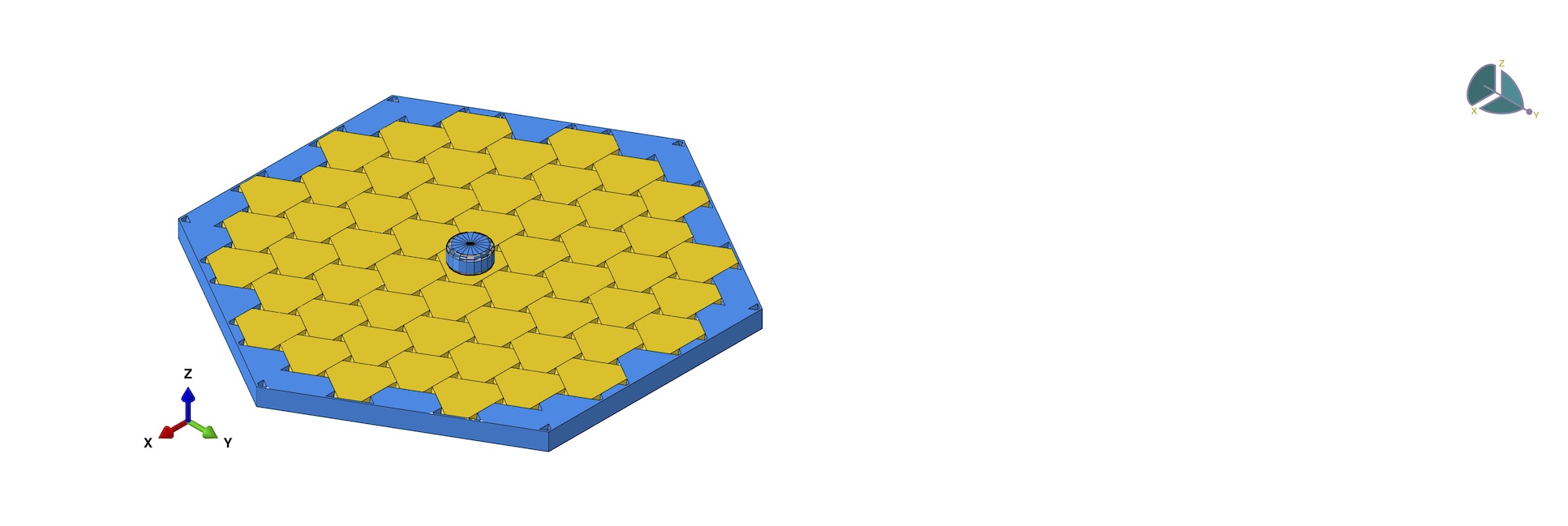}
        \label{36-A-1_assembly}}
    \subfigure[]{
        \assemblyPlotA{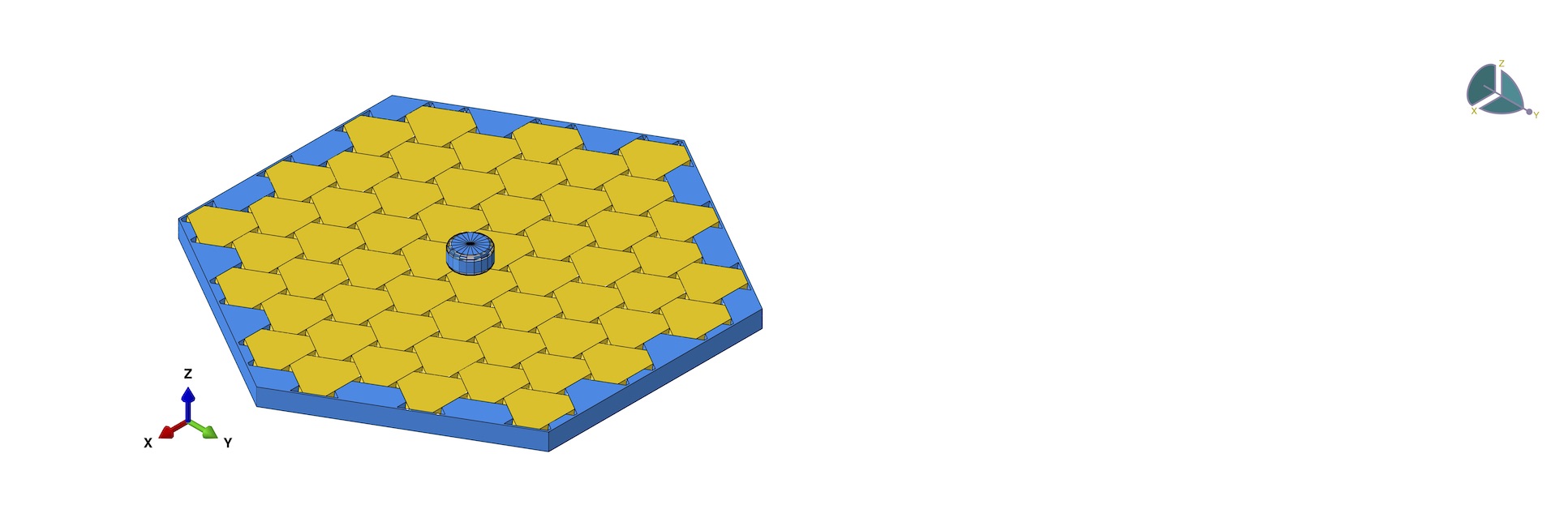}
        \label{36-B-1_assembly}}
    \caption{Assembly configurations: (a) $[3^6]$-A, (b) $[3^6]$-B.}
    \label{36_assembly}
\end{figure}

\begin{figure}[ht]
    \centering
    \subfigure[]{
        \assemblyPlotA{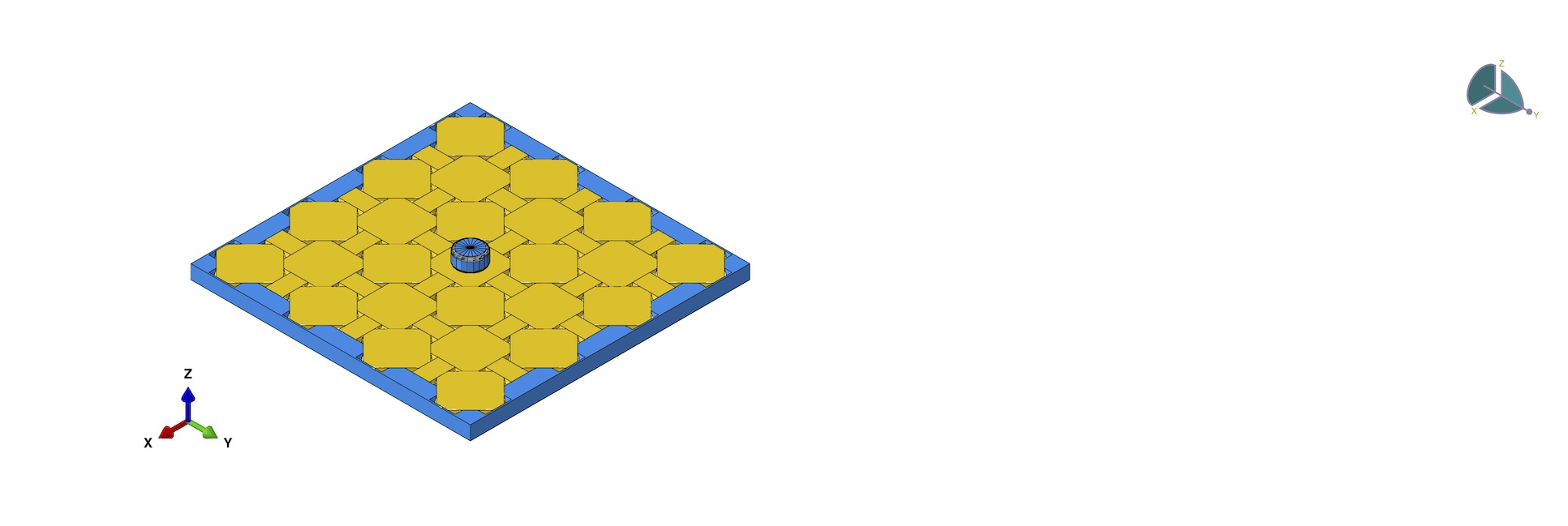}
        \label{482-A-1_assembly}}
    \subfigure[]{
        \assemblyPlotA{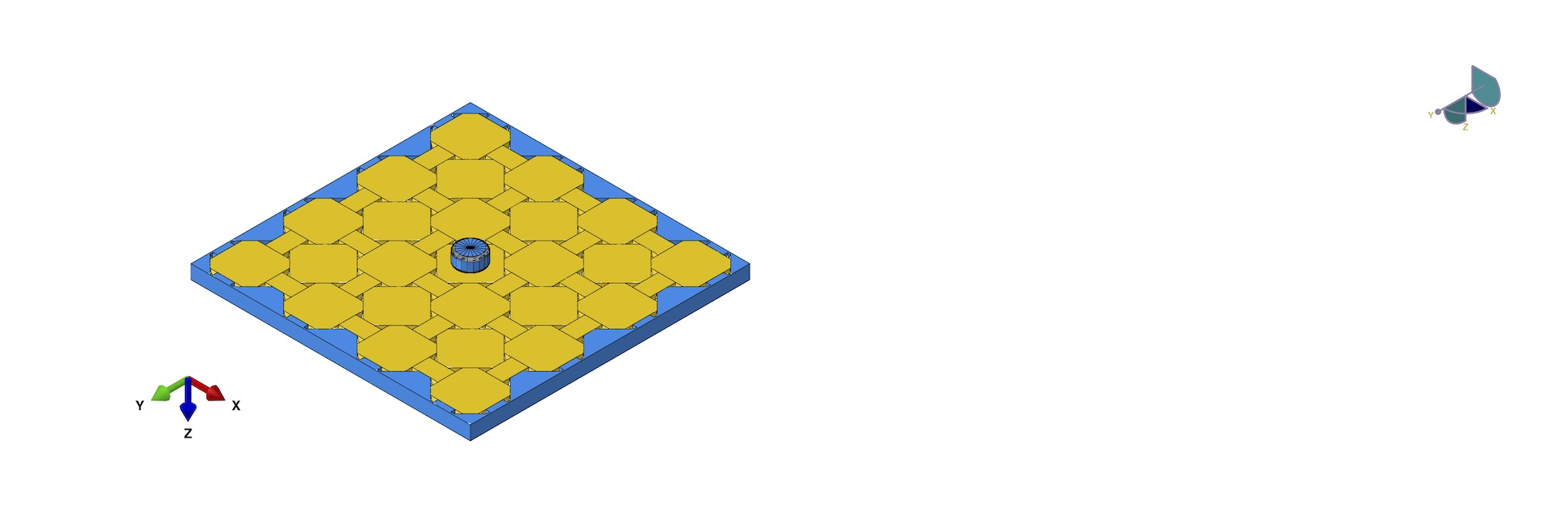}
        \label{482-A-2_assembly}}
    \subfigure[]{
        \assemblyPlotA{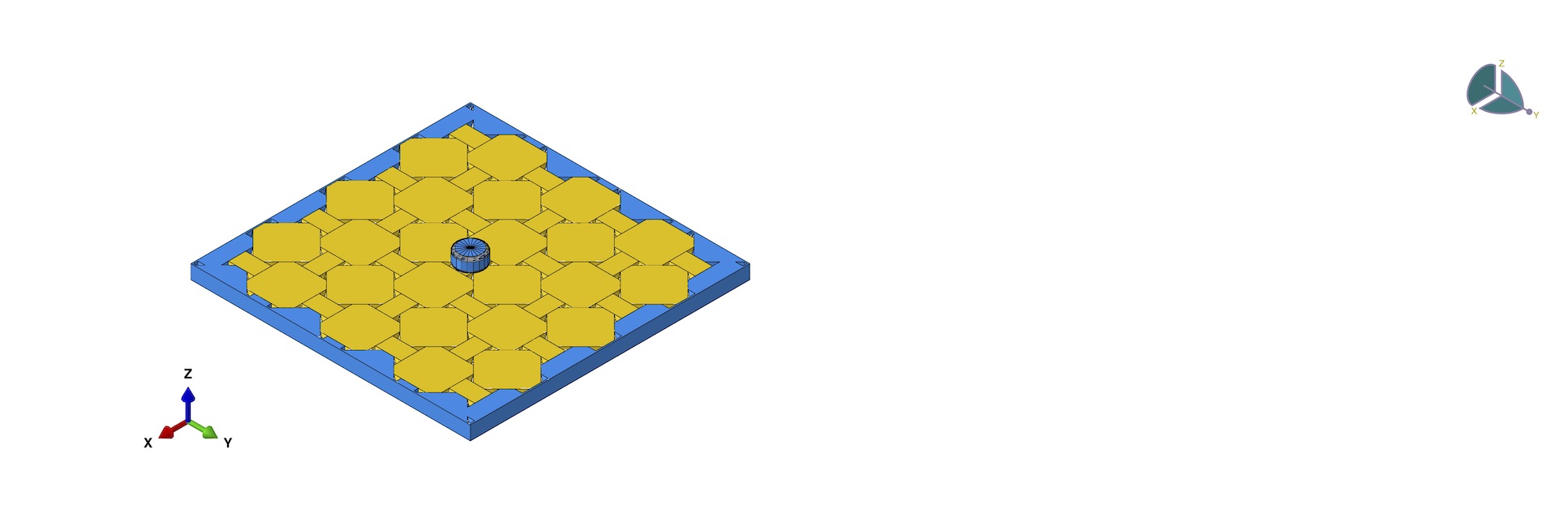}
        \label{482-B-1_assembly}}
    \caption{Assembly configurations: (a) $(4.8^2)$-A(-), (b) $(4.8^2)$-A(+), (c) $(4.8^2)$-B.}
    \label{482_assembly}
\end{figure}

\begin{figure}[ht]
    \centering
    \subfigure[]{
        \assemblyPlotA{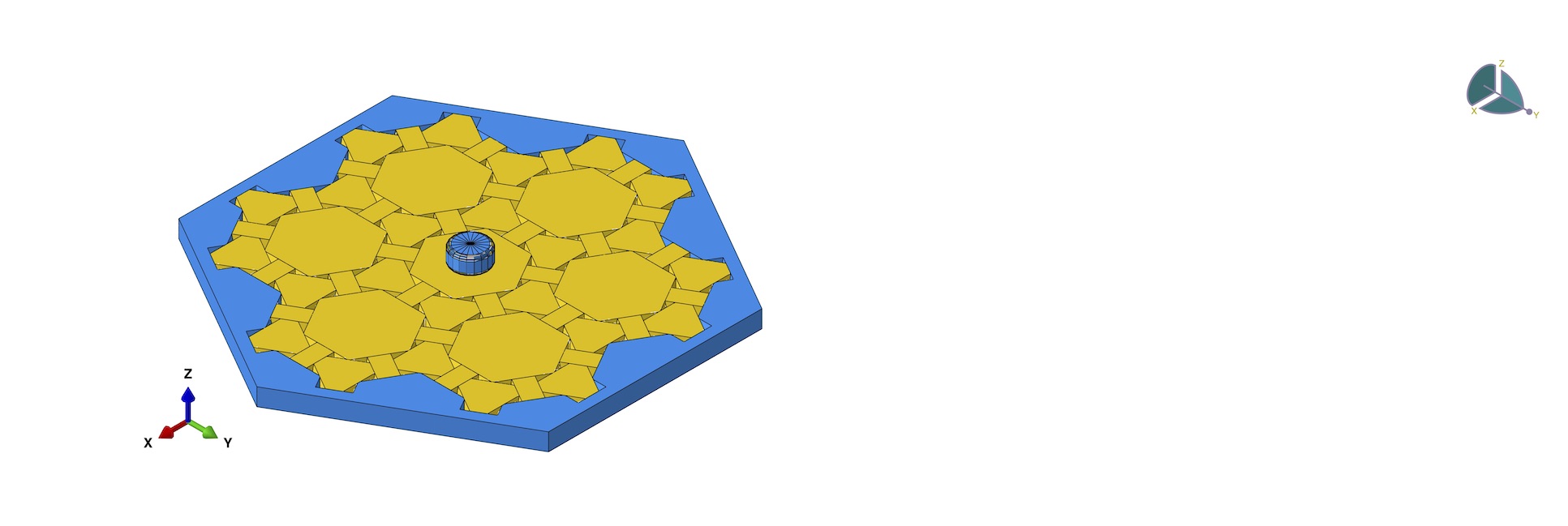}
        \label{4612-A-1_assembly}}
    \subfigure[]{
        \assemblyPlotA{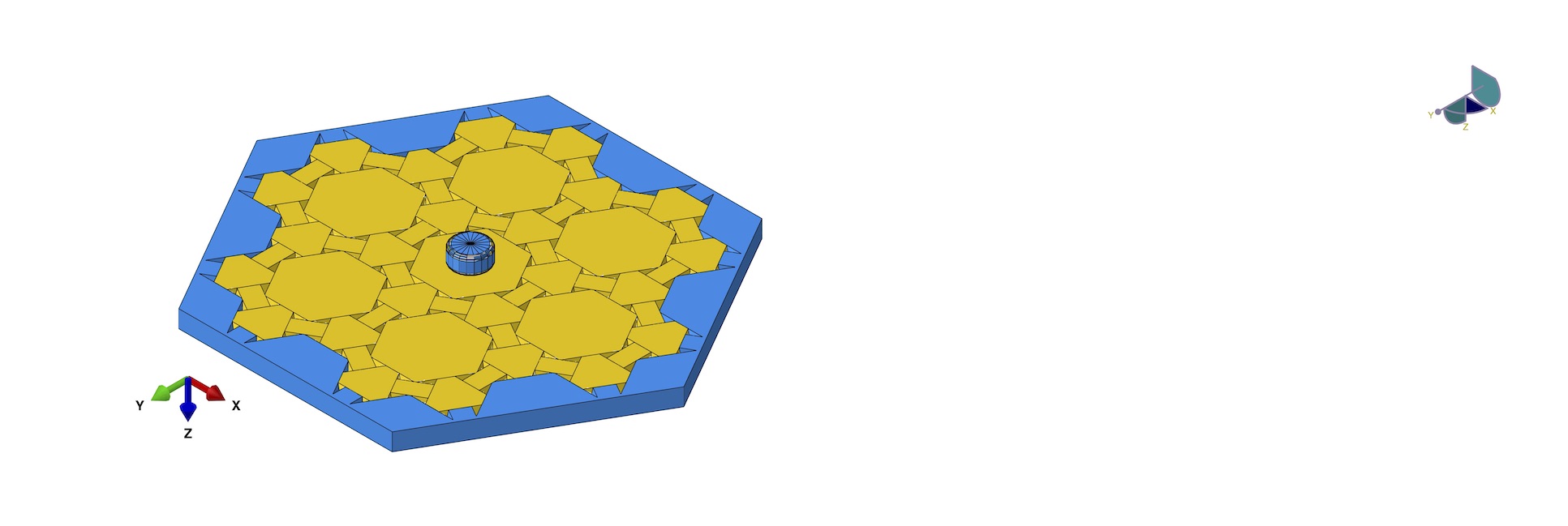}
        \label{4612-A-2_assembly}}
    \subfigure[]{
        \assemblyPlotA{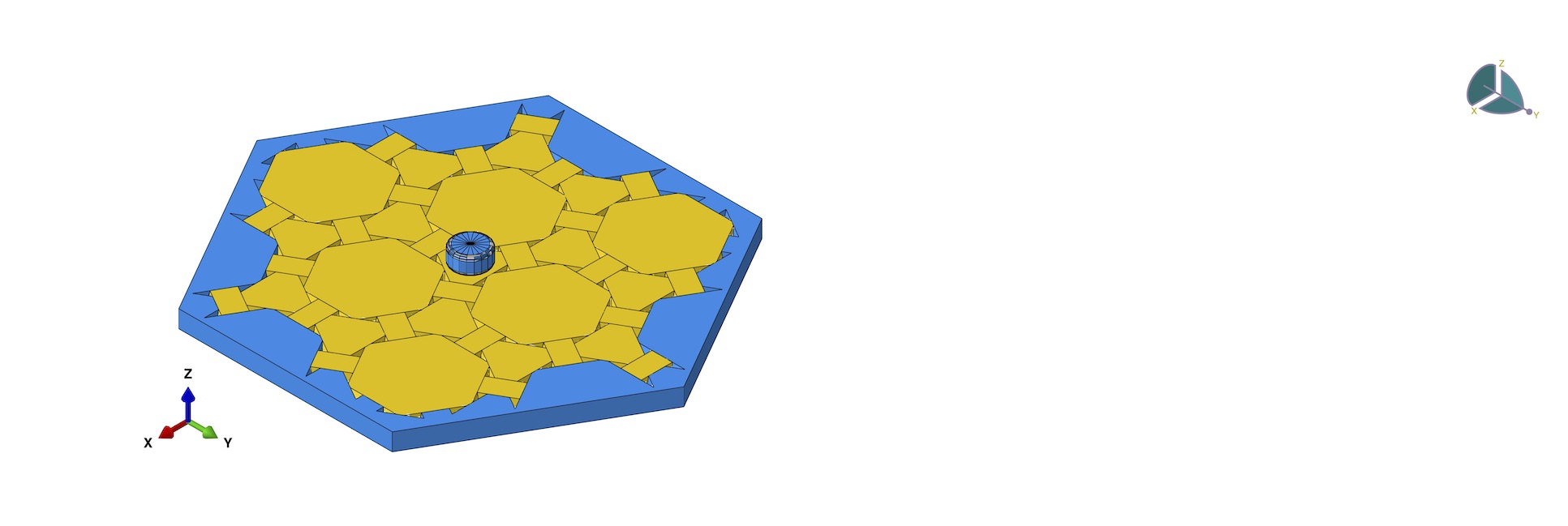}
        \label{4612-B-1_assembly}}
    \subfigure[]{
        \assemblyPlotA{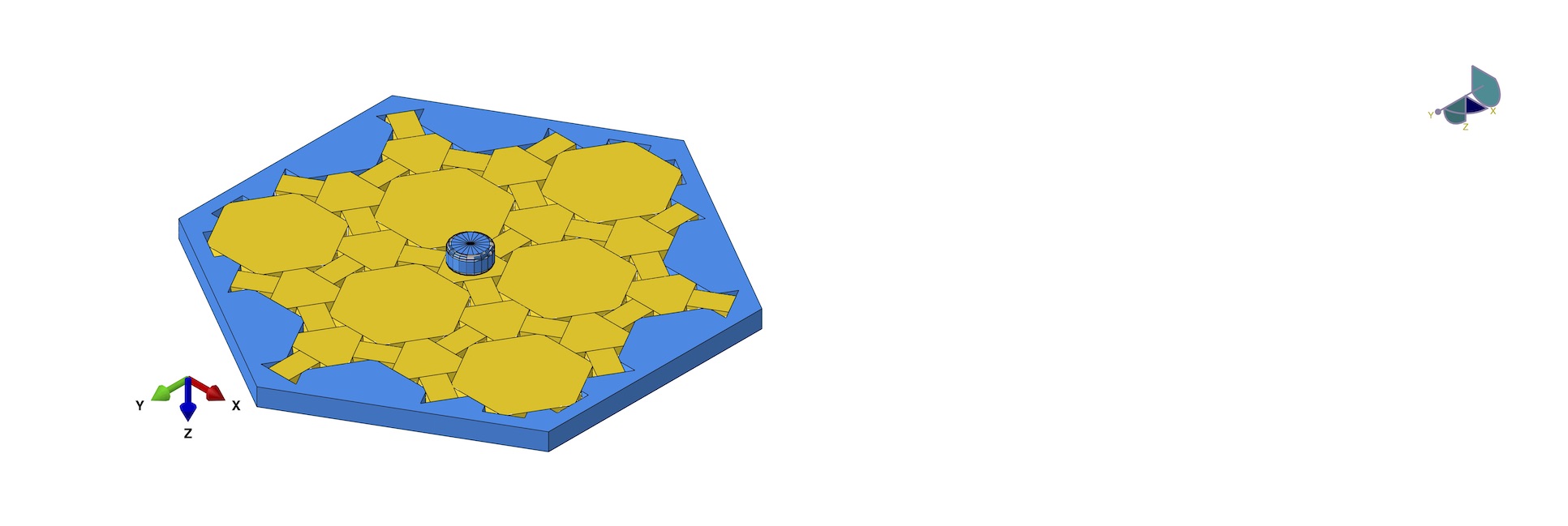}
        \label{4612-B-2_assembly}}
    \subfigure[]{
        \assemblyPlotA{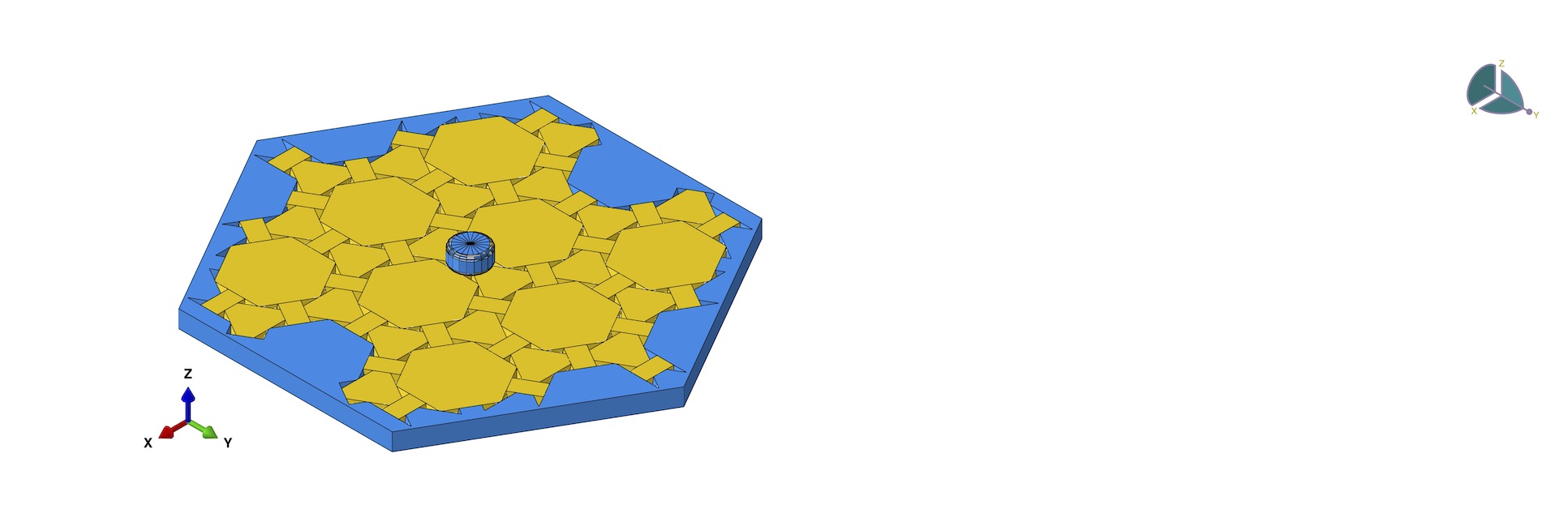}
        \label{4612-C-1_assembly}}
    \subfigure[]{
        \assemblyPlotA{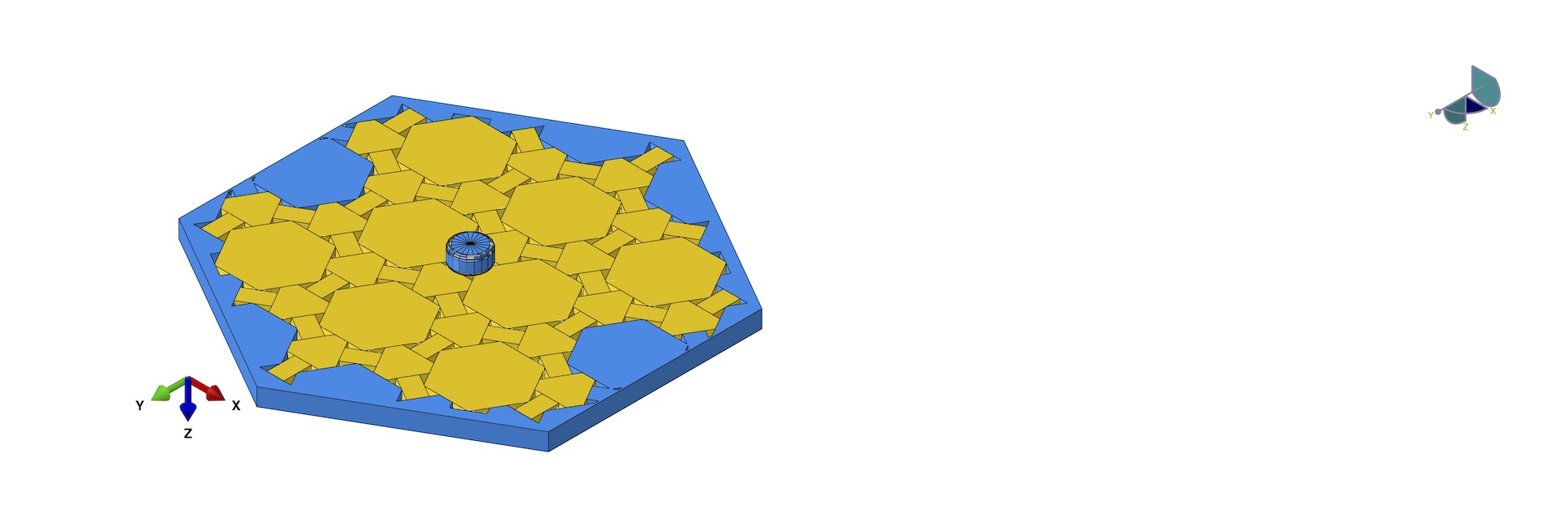}
        \label{4612-C-2_assembly}}
    \caption{Assembly configurations: (a) (4.6.12)-A(-), (b) (4.6.12)-A(+), (c) (4.6.12)-B(-), (d) (4.6.12)-B(+), (e) (4.6.12)-C(-), (f) (4.6.12)-C(+).}
    \label{4612_assembly}
\end{figure}
\newpage

\FloatBarrier
\section{Analysis Approach}
All model configurations are analyzed within the context of the finite element analysis. Finite element models were computed with an explicit solver (ABAQUS) to obtain the load - displacement response of all TIM configurations under quasi-static transverse loading from point load under displacement control conditions. The computed reaction forces were filtered using a second order Butterworth filter with a cutoff frequency of 50 Hz.

The frame was made to be undeformable and fixed in space.
The elastic modulus of the unit elements was assigned to be $E = 1.83$ GPa, the Poisson ratio $\nu = 0.35$. These properties are motivated by a 3D printing manufacturing approach for the physical realization of interlocked assemblies~\cite{Siegmund2016ManufactureAssemblies}. Contact was defined between all bodies with a stiff linear pressure-overclosure relationship and a coefficient of friction of $\mu = 0.2$. A density $\rho = 0.95$ g/cm$^3$ was considered and mass scaling by a factor of 100 was employed to reduce computation time.
8-node reduced integration hexahedral elements (C3D8R) were used to mesh the blocks, while solid 4-node tetrahedral elements (C3D4) were used for the rigid frame. Enhanced hourglass control was used on on the hexahedral elements to reduce an observed tendency for hourglassing with default hourglass control.

Mesh convergence was evaluated by comparing force-deflection data for all models with mesh seed size over a range from 0.15 $H_0$ to 0.21$H_0$ in increments of 0.01$H_0$. It was found that convergent results for the computed force-deflection behavior were obtained for almost all cases if the mesh seed size is 0.16$H_0$. The exception to that finding were the $[4^4]$-A and $[4^4]$-B configurations. These cases were susceptible to a perfect alignment of meshes across contacts which tends to result in a nontraditional hourglassing across contact interfaces. Such cross-contact hourglassing would create a local interlocking feature between the blocks and prevent sliding. A seed size of 0.17$H_0$ was used for the  $[4^4]$-A and $[4^4]$-B configurations to avoid the mesh alignment issue.

\newpage


\end{document}